\documentclass[10pt,aps,prb,amsmath,twocolumn,amssymb,floatfix,showpacs,nofootinbib,longbibliography]{revtex4-2}
\usepackage{float}
\usepackage{braket}
\usepackage[dvipsnames]{xcolor}
\usepackage{subfigure}
\usepackage[colorlinks=true,linktoc=page,linkcolor=blue,citecolor=blue,urlcolor=blue]{hyperref}
\usepackage{orcidlink}
\usepackage[normalem]{ulem}
\newcommand{\vect}[1]{\boldsymbol{\mathrm{#1}}}
\newcommand{\Eq}[1]{Eq.~(\ref{#1})}
\newcommand{\Fig}[1]{Fig.~\ref{#1}}
\mathchardef\mhyphen="2D % Define a "math hyphen"

\allowdisplaybreaks
\begin{document}
%=============START of MAIN PAPER===============

\title{Laser-induced topological phases in monolayer amorphous carbon}
\author{Arnob Kumar Ghosh~\orcidlink{0000-0003-0990-8341}}
\email{arnob.ghosh@physics.uu.se}
\affiliation{Department of Physics and Astronomy, Uppsala University, Box 516, 75120 Uppsala, Sweden}

\author{Quentin Marsal~\orcidlink{0000-0001-6506-1582}}
\email{quentin.marsal@physics.uu.se}
\affiliation{Department of Physics and Astronomy, Uppsala University, Box 516, 75120 Uppsala, Sweden}

\author{Annica M. Black-Schaffer~\orcidlink{0000-0002-4726-5247}}
\email{annica.black-schaffer@physics.uu.se}
\affiliation{Department of Physics and Astronomy, Uppsala University, Box 516, 75120 Uppsala, Sweden}

%--------------------------------------------------------
%--------------------------------------------------------
\begin{abstract}
Driving non-topological materials out of equilibrium using time-periodic perturbations, such as circularly-polarized laser light, is a compelling way to engineer topological phases. At the same time, topology has traditionally only been considered for crystalline materials.
Here we propose an experimentally feasible way of driving monolayer amorphous carbon topological.
We show that circularly polarized laser light induces both regular and anomalous edge modes at quasienergies $0$ and $\pm \pi$, respectively. We also obtain a complete topological characterization using an energy- and space-resolved topological marker based on the spectral localizer. 
Additionally, by introducing atomic coordination defects in the amorphous carbon, we establish the importance of the local atomic coordination in topological amorphous materials. Our work establishes amorphous systems, including carbon, as a versatile and abundant playground to engineer topological phases.
\end{abstract}
%--------------------------------------------------------
%--------------------------------------------------------

\maketitle

The Haldane model describing the anomalous quantum Hall effect~(AQHE) in a hexagonal lattice without requiring an external magnetic field has been a foundation for topological phases of matter~\cite{HaldanePRL1988, qi2011topological, BernevigBOOK}. 
However, realizing the Haldane model in a static solid-state system has been experimentally challenging~\cite{ChangQAHRMP2023}. 
Moreover, for the best advantage of topological properties, it is preferable to have systems where topology can be controlled using a simple external control knob. 

Floquet engineering via a time-periodic modulation of the Hamiltonian is precisely a simple and efficient way of manipulating band topology in an otherwise topologically trivial material~\cite{Kitagawacharacterization2010,kitagawa11transport,lindner11floquet, FloquetGuPRL2011, Rudner2013, Usaj2014, PerezPRA2015, Eckardt2017,oka2019, NHLindner2020, Bao2022, GhoshJPCMReview2024}, resulting in topological edge states at zero and/or finite energy in the spectrum~\cite{Kitagawacharacterization2010, JiangColdAtomPRL2011, Rudner2013, Piskunow2014, Usaj2014, PerezPRA2015, Yan2017, Eckardt2017, NHLindner2020}. For example, the Haldane model and its AQHE have recently been experimentally indirectly observed in monolayer graphene driven by a circularly polarized laser~\cite{McIver2020, lesko2025opticalcontrolelectronsfloquet}. Floquet topological phases have also been observed in other experimental setups, such as photonic systems~\cite{RechtsmanExperiment2013, Maczewsky2017}, acoustic setups~\cite{Peng2016, fleury2016floquet}, and cold-atomic systems~\cite{Jotzu2014, Wintersperger2020}.

While taking into account the spatial symmetries of crystalline materials greatly simplifies the characterization of their topology~\cite{FuKane2007, PoNatCommun2017, SlagerPRX2017, bradlyn2017}, this should not overshadow the fact that those symmetries are not required, as a recent work has emphasized in amorphous materials~\cite{CorbaeEPL2023}.
Importantly, amorphous solids are both abundant in nature and can be obtained in a vast diversity of materials, thus drastically extending the possibilities for topological physics. 
While amorphous materials lack translation symmetry, they are still characterized by a well-defined local atomic environment~\cite{zallen_physics_1998}, set by the chemistry of their atomic constituents.
This local order is a key feature for topology~\cite{AgarwalaPRL2017,MitchellNP2018,PoyhonenNC2018,SahlbergPRR2020,YangAMPRL2019,CostaNL2019,MarsalPNAS2020,ZhouLSA2020,WangPRLStructural2021,CorbaePRB2021,LiPRLTyd2021,WangAMIPRL2022} and amorphousness enhances the diversity of achievable local environments~\cite{legallo2020}.
Experimental observations in mechanical~\cite{MitchellNP2018}, photonic~\cite{Liu2020,ZhouLSA2020,Jia2022}, and condensed matter platforms~\cite{CorbaeNatMater2023,CiocysNatComm2024} further confirm the role played by local order.

Being able to combine the precise control of topology offered by a laser drive with the enhanced freedom of atomic structures in amorphous materials would open the way to an abundance of material platforms hosting controllable topologically protected modes not restricted by crystallinity.
In this work, we demonstrate that a circularly polarized laser induces topological phases in monolayer amorphous carbon, an amorphous version of graphene~\cite{TohNature2020}. We find both regular and anomalous edge modes at quasienergies 0 and $\pm \pi$, respectively. To overcome the broken translation invariance and the periodicity of the energy spectrum, we characterize its topology based on the spectral localizer~\cite{GhoshSL2024}.
We also study substitutional defects~\cite{KrasheninnikovPRL2009,ZhouPRL2012,WangNL2012, RamasseNL2013,YanCSR2018} introducing fourfold-coordinated sites in otherwise threefold-coordinated monolayer amorphous carbon. We find topology to be robust to a moderate amount of such defects, but destroyed by a global change of the local atomic structure. This establishes local coordination as a key tool for topology in amorphous materials.
Our results demonstrate that monolayer amorphous carbon becomes topological, despite lacking both the long-range order and bipartiteness of graphene, and highlight amorphous materials as versatile playgrounds for topological phases.

%~~~~~~~~~~~~~~~~~~~~~~~~~~~~~~~~~~~~~~~~~~~~~~~~~~~~~~~~~
%~~~~~~~~~~~~~~~~~~~~~~~~~~~~~~~~~~~~~~~~~~~~~~~~~~~~~~~~~
\begin{figure}
	\centering
	\includegraphics[width=0.3\textwidth]{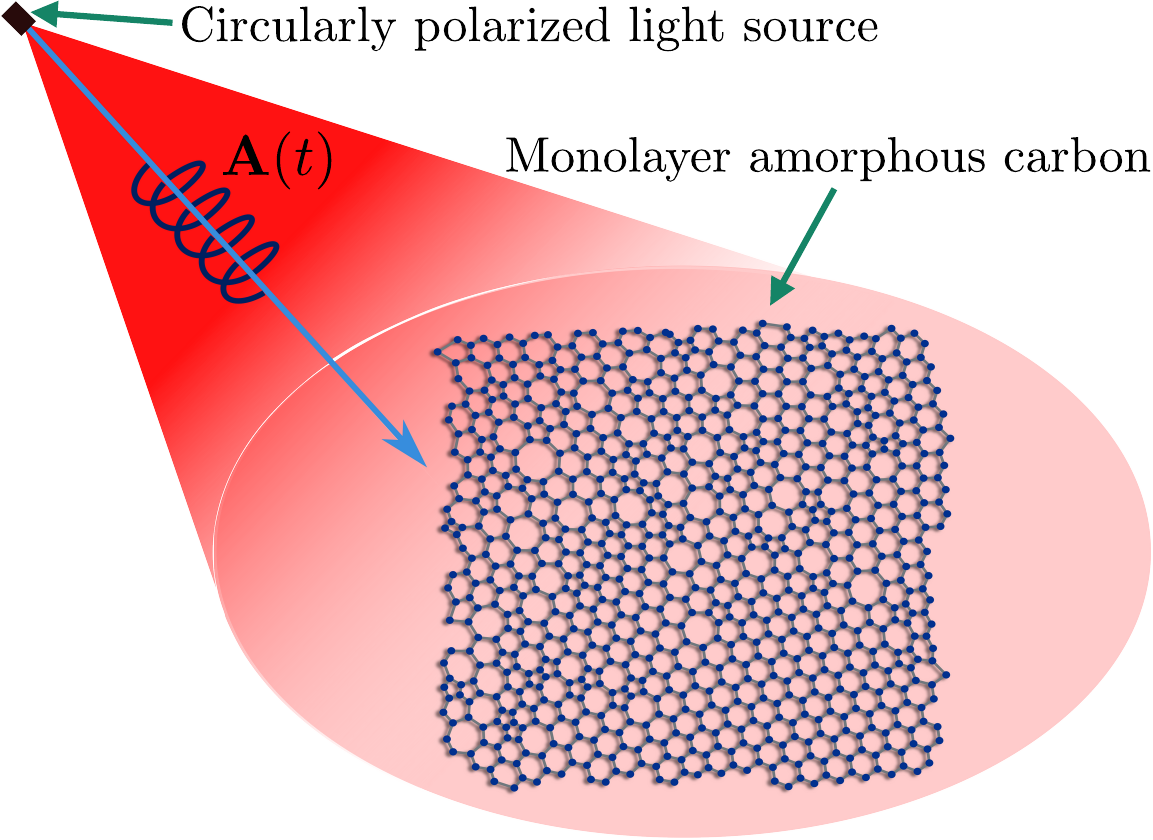}
	\caption{Schematic representation of a circularly polarized laser beam with field $\vect{A}(t)$ on a monolayer amorphous carbon to engineer topological phases.
	}
	\label{Fig:Schematics}
\end{figure}
%~~~~~~~~~~~~~~~~~~~~~~~~~~~~~~~~~~~~~~~~~~~~~~~~~~~~~~~~~
%~~~~~~~~~~~~~~~~~~~~~~~~~~~~~~~~~~~~~~~~~~~~~~~~~~~~~~~~~

\textcolor{blue}{\textit{Material.---}}
We consider monolayer amorphous carbon~\cite{TohNature2020}, which can be experimentally obtained by electron beam irradiation of graphene~\cite{KotakoskiPRL2011}. This locally disorders the polygons formed by the carbon atoms, see lattice in Fig.~\ref{Fig:Schematics}, while crystalline graphene consists of only hexagons. 
Chemical vapor deposition at high temperatures has also been employed to create different polygons in otherwise monolayer graphene~\cite{HuangNP2011,KwanpyoACSNano2011,YazyevNN2014,TohNature2020}. 
Due to the $sp^2$-hybridization of the carbon atoms, monolayer amorphous carbon remains exactly threefold-coordinated~\cite{TohNature2020} and, despite fluctuations resulting from geometric constraints, the average length $d_0$ of nearest-neighbor bonds is also similar to graphene. Such well-defined local properties are a key feature of amorphous materials~\cite{zallen_physics_1998}.
In the last part, we additionally introduce coordination defects in the form of fourfold-coordinated sites. Such defects have been experimentally realized by bombarding graphene to create vacancies and substitutional transition metal or silicon sites~\cite{KrasheninnikovPRL2009, ZhouPRL2012, WangNL2012, RamasseNL2013, YanCSR2018}.

The preserved local structure of monolayer amorphous carbon allows us to model the electronic properties
using a spinless tight-binding Hamiltonian for the $p_z$ orbital on each carbon atom with nearest-neighbor hopping amplitude $t_h$,
\begin{equation}
    H = \sum_{\langle i,j \rangle} t^{ij}_h  c_i^\dagger c_j + {\rm H.c.} ,
    \label{Eq:Ham}
\end{equation} 
set on an amorphous lattice with exact threefold coordination, see Fig.~\ref{Fig:Schematics}. 
The lattice is obtained using a Voronoi tesselation of a random set of points following Ref.~\cite{MarsalPNAS2020}. 
We incorporate the effect of varying bond lengths in the lattice through an exponential dependence of the hopping parameter, $t^{ij}_h= t_h \exp\left[-\left( \lvert \vect{r}_j - \vect{r}_i \rvert -d_0\right)/\lambda \right]$, where we set $t_h = 1$ as the energy unit, $d_0=1/\sqrt{3}$, and $\lambda$ encodes the decay length of the hopping integral. 
We set $\lambda=\sqrt{3}d_0$, corresponding to the graphene second-nearest-neighbor distance, but our results do not depend on the choice of $\lambda$, as long as it remains of the order of $d_0$.

\textcolor{blue}{\textit{Drive.---}}
To induce topology in a monolayer amorphous carbon, we consider a driving protocol consisting of a circularly-polarized laser, represented by a vector potential $\vect{A}(t)=A(\cos \Omega t, \sin \Omega t)$, where $\Omega$ is the frequency, $T = \frac{2\pi}{\Omega}$ its period, and $A$ proportional to the light intensity. 
We consider the beam spot to be much larger than the system's dimension, such that there is no spatial variation in the light intensity, see~\Fig{Fig:Schematics}. 
The circular polarization breaks time-reversal symmetry and is known to generate an energy gap in graphene, thereby allowing topological phases~\cite{Usaj2014,Piskunow2014}. 
The driving field $\vect{A}(t)$ enters the Hamiltonian $H$ in \Eq{Eq:Ham} through Peierls phase substitution in the electron hopping, such that $t_h^{ij} \rightarrow t_h \exp \left[-  i \int_{\vect{r}_i}^{\vect{r}_j} \vect{A}(t) \cdot d\vect{r} -\left( \lvert \vect{r}_j - \vect{r}_i \rvert -d_0\right)/\lambda\right]$. 

The time-periodic Hamiltonian $H(t)=H(t+T)$ can be treated using Floquet theory~\cite{FLoquetpaper1883, Eckardt2017, oka2019, NHLindner2020, Bao2022, GhoshJPCMReview2024}, allowing an effective time-independent description of the system. We compute the Floquet operator $U(T,0)$ from the time-dependent Hamiltonian in a time-ordered~(TO) notation as $U(T,0)= {\rm TO} \exp \left[ -i \int_{0}^{T} dt H(t) \right] $.
Numerically, we compute $U(T,0)$ by a Trotterization process as $U(T,0)=\prod_{j=0}^{N-1} U(t_{j}+\delta t,t_j)$, where $U(t_{j}+\delta t,t_j)=e^{-i H(t_j)\delta t}$, $\delta t = \frac{T}{N}$, and $t_j=j \delta t$, using $200$ Trotter steps. The effective time-independent Floquet Hamiltonian is then computed as $H_F=-i \ln [U(T,0)]$, which captures the stroboscopic dynamics of the system and determines its spectral properties, including the local and total density of states~(LDOS and TDOS). The spectrum of $H_F$ is $\Omega$-periodic, such that the spectrum lies in the range $(-\Omega/2,\Omega/2]$ with $\pm \Omega/2 \equiv \pm \pi$. Thus, the driven system can host topological edge states at gaps at both $0$- and $\pm \pi$ quasienergies~\cite{Kitagawacharacterization2010, JiangColdAtomPRL2011, Rudner2013,Piskunow2014,Usaj2014,PerezPRA2015,Yan2017,Eckardt2017,NHLindner2020}.

%~~~~~~~~~~~~~~~~~~~~~~~~~~~~~~~~~~~~~~~~~~~~~~~~~~~~~~~~~
%~~~~~~~~~~~~~~~~~~~~~~~~~~~~~~~~~~~~~~~~~~~~~~~~~~~~~~~~~
\begin{figure}
	\centering
	\subfigure{\includegraphics[width=0.49\textwidth]{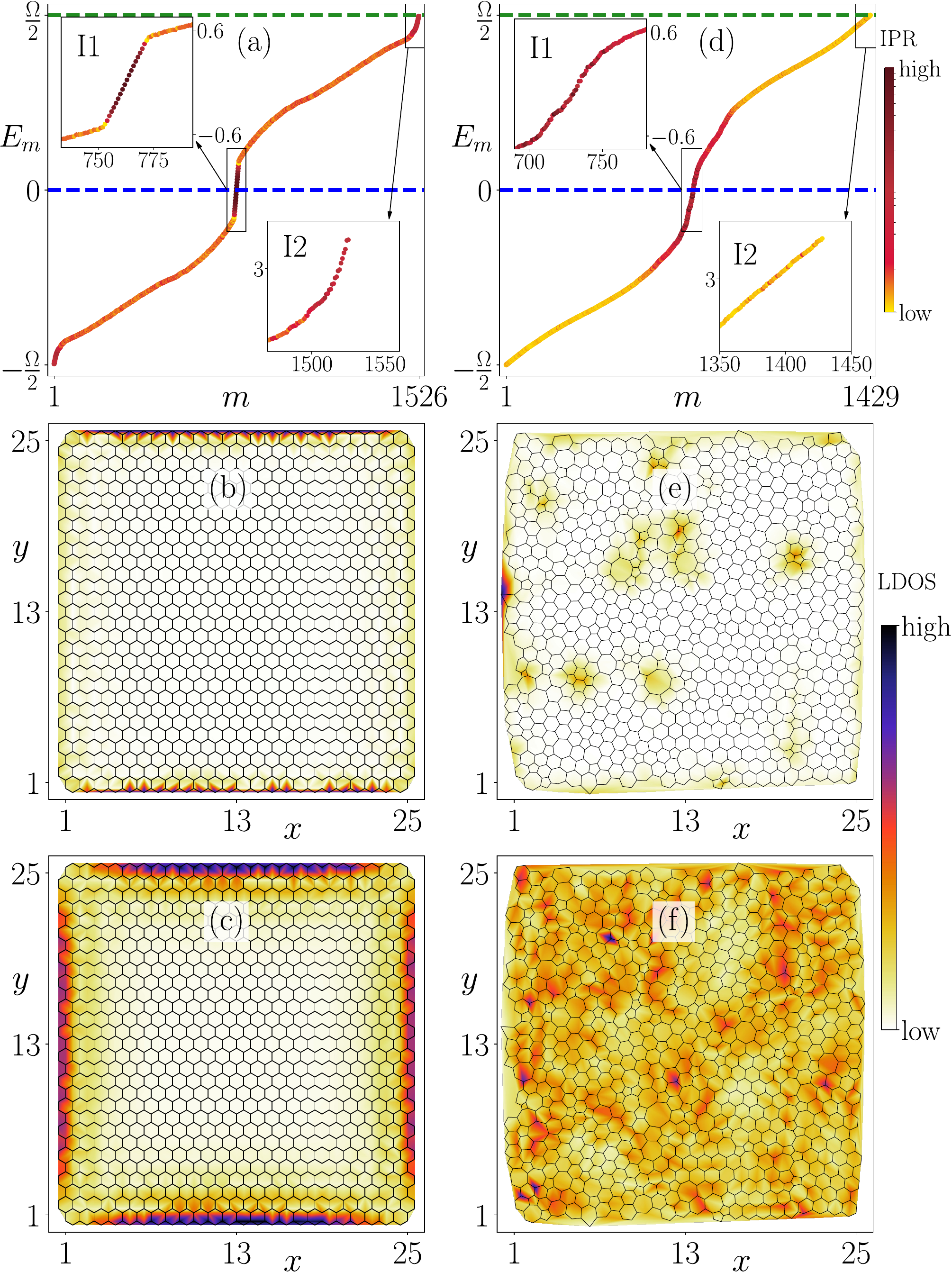}}
	\caption{(a) Quasienergy spectrum $E_m$ as a function of eigenstate index $m$ for crystalline graphene. Color encodes the IPR of the given state. Insets I1 and I2 show zoomed-in spectra close to $E=0$ and $\pi$, respectively. (b,c) Spatially resolved LDOS associated with states at quasienergy gaps $0$ and $\pi$, respectively. (d-f) Repeats (a-c) but for monolayer amorphous carbon. Here $A=2.5$, $\Omega=2.0$. For LDOS computation, we use a quasienergy window of $[-0.3,0.3]$ in $0$-gap and $[\pm \Omega/2, \mp 0.1]$ in the $\pi$-gap. Number of atoms in (a-c) is $1526$ and in (d-f) is $1442$. 
	}
	\label{Fig:Polycrystalline}
\end{figure}
%~~~~~~~~~~~~~~~~~~~~~~~~~~~~~~~~~~~~~~~~~~~~~~~~~~~~~~~~~
%~~~~~~~~~~~~~~~~~~~~~~~~~~~~~~~~~~~~~~~~~~~~~~~~~~~~~~~~~

\begin{figure}
	\centering
	\subfigure{\includegraphics[width=0.48\textwidth]{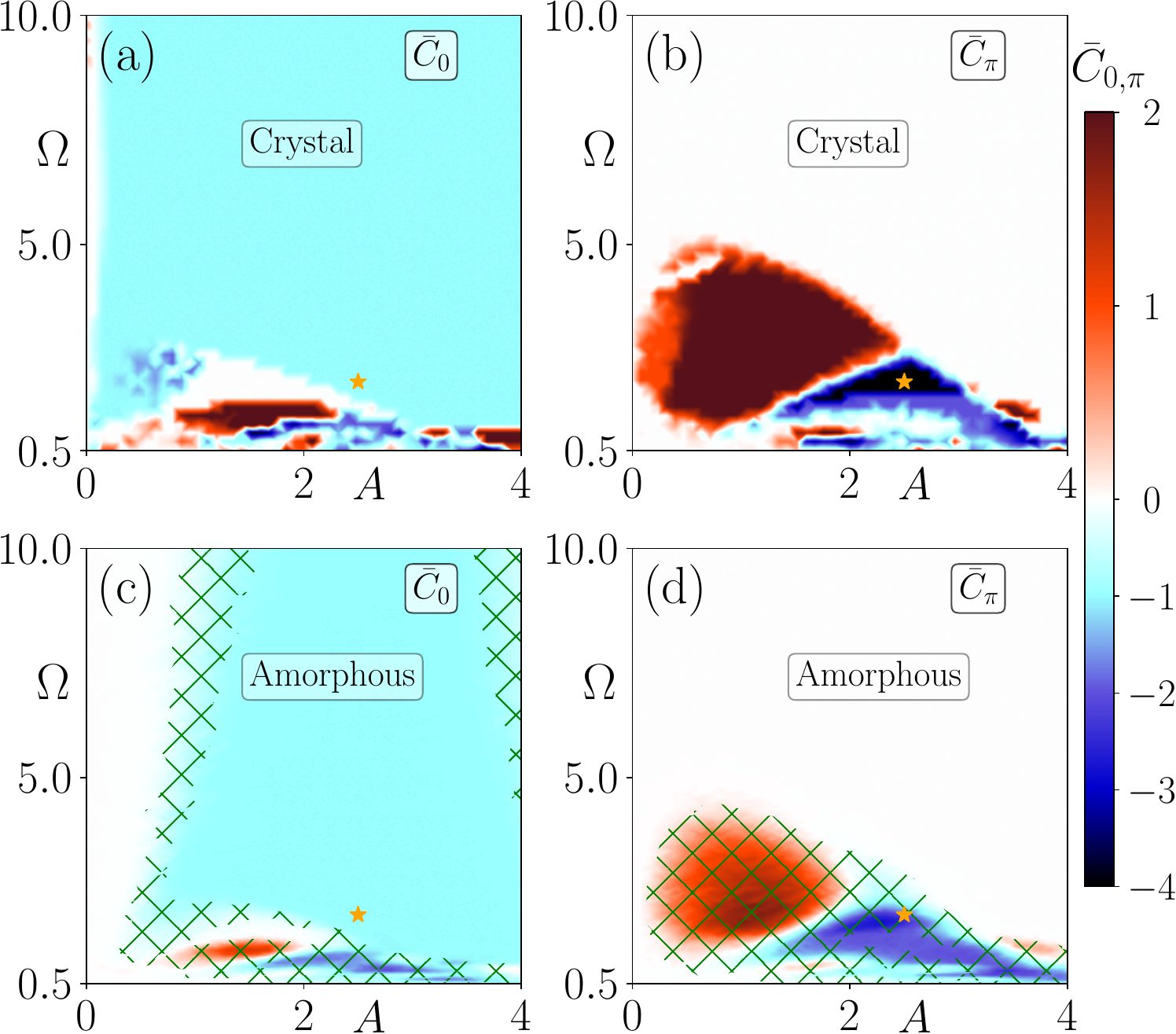}}
	\caption{Average Chern numbers (a) $\bar{C}_0$ and (b) $\bar{C}_\pi$ as a function of driving amplitude $A$ and frequency $\Omega$ for crystalline graphene. Number of atoms in (a-b) is $880$. (c-d), Repeats (a-b) for monolayer amorphous carbon. Green lines indicate regions with a finite variance of the Chern numbers. Orange star marks parameters used in Fig.~\ref{Fig:Polycrystalline}. Averages of $C_{x,y,E}$ in each sample are calculated using a $7 \times 7$ square in the center of the system, and $30$ different amorphous configurations are sampled with a system consisting of $837$ atoms.
	}
	\label{Fig:PhaseDiagramAmorphousGraphene}
\end{figure}

\textcolor{blue}{\textit{Driven topological phases.---}}
We first demonstrate that the circularly polarized laser induces topological edge states through the spectral properties of driven monolayer amorphous carbon in Fig.~\ref{Fig:Polycrystalline}. For this, we choose driving parameters such that the system hosts edge modes in both the $0$- and $\pi$-gaps (see orange star in Fig.~\ref{Fig:PhaseDiagramAmorphousGraphene}). 
Before considering an amorphous system, we study (crystalline) graphene. 
We plot the quasienergy spectrum $E_m$ as a function of the eigenvalue index $m$ in Fig.~\ref{Fig:Polycrystalline}(a) for driven graphene, with zoom-ins for the $0$ and $\pi$-gaps. We also compute the inverse participation ratio~(IPR) of the eigenstates: ${\rm IPR}(\ket{\psi_m}) = \sum_i\left|\braket{i|\psi_m}\right|^4$ and color-code each eigenstate by its IPR. 
The IPR indicates localization by probing the inverse number of sites occupied by the state $\ket{\psi_m}$, ranging from $0$ for a delocalized state in an infinite system, to $1$ for a fully localized state. The IPR thus distinguishes between in-gap  `localized' edge states (dark red) and bulk states (yellow to light red) in Fig.~\ref{Fig:Polycrystalline}(a), with the color-code also revealing the gap sizes.
Next, we compute the LDOS for the states in the $0$- and $\pi$-gaps in Fig.~\ref{Fig:Polycrystalline}(b,c), respectively. 
The LDOS shows that the gap states are localized along the edges, indicating topological origin.
The presence of drive-induced topological modes in graphene is consistent with previous studies~\cite{Usaj2014,Piskunow2014,MikamiBW2016,McIver2020,MerboldtNP2025}.

Next, we turn to monolayer amorphous carbon. 
Figure~\ref{Fig:Polycrystalline}(d) and inset I1 surprisingly shows a drive-induced gap opening at $E=0$, just as in graphene, with clearly localized in-gap states. The LDOS Fig.~\ref{Fig:Polycrystalline}(e) shows edge localization of this $0$-mode. We also observe some localized states in the bulk, which are defect states appearing in amorphous systems and resulting in a finite bulk LDOS, also in the gap.
For the $\pi$-gap, the IPR reveals only a few localized states, indicating a gap-closing at $\pm \pi$ compared to graphene, due to disorder-induced states. The LDOS for the $\pi$-modes in Fig.~\ref{Fig:Polycrystalline}(f) also does not exhibit a clear signature of the edge-localization. Thus, the $\pi$-modes are mostly unstable in monolayer amorphous carbon, most likely due to a smaller spectral gap protecting them from the bulk states. We also discuss this in terms of the stability of the topological invariant in Fig.~\ref{Fig:PhaseDiagramAmorphousGraphene}.
However, we can still find some parameters for which $\pi$-modes are stable, see the supplemental material~(SM)~\cite{supp}.
These results showcase that amorphous materials can be driven into topological phases, with associated edge modes. Notably, in terms of monolayer amorphous carbon, this is despite losing both the long-range order and bipartiteness of crystalline graphene. 

\textcolor{blue}{\textit{Topological characterization.---}}
Having established the existence of edge states, we provide a complete topological characterization of the system. This is challenging since monolayer amorphous carbon layer breaks translational symmetry and the driven system hosts modes at $0$- and $\pi$-quasienergies. Thus, we need both a real-space and energy-resolved topological invariant to fully characterize the  system~\cite{TitumPRX2016,GhoshSL2024,GavenskyPRX2025}. For this, we utilize the spectral localizer to obtain a space- and energy-resolved topological index~\cite{LoringAnnPhys2015, loring2017finitevolume,loring2019guide,GhoshSL2024}. 
The spectral localizer operator includes the system's position operators $X$ and $Y$ in the $x$- and $y$-directions, respectively, and the Hamiltonian, which for a driven system is taken as the effective Floquet Hamiltonian $H_F$~\cite{GhoshSL2024}. Explicitly, the spectral localizer is defined as~\cite{LoringAnnPhys2015, loring2017finitevolume,loring2019guide,GhoshSL2024}
\begin{align}
    L_{x,y, E} \! \left(X,Y, H_{F}\right)\!=\!\kappa \left[ (X\!\! -\!x I) \tau_x \!+\!  (Y\! \!-\!y I)  \tau_y\right] \!+ \!(H_{F} \!-\! EI)  \tau_z.
    \label{localizer2D}
\end{align}
Here $x,y$, and $E$ define the positions and energy where the spectral localizer is computed, $\kappa$ ensures a compatible weight between the position operators and the Hamiltonian, and the Pauli matrices $\vect{\tau}$ satisfies the anti-commutation relation $\{ \tau_i,\tau_j\}=2\delta_{i,j}$, forming the Clifford representation. We use $\kappa=0.01$, which ensures stable evaluation.
Utilizing $L_{x,y, E}\left(X,Y, H_{F}\right)$, the topological index, or Chern number, is defined as~\cite{LoringAnnPhys2015,GhoshSL2024}
\begin{align}
  C_{x,y,E} = \frac{1}{2}{\rm sig}\left[L_{x,y, E} \left(X,Y, H_F\right)\right], 
     \label{localChern}
\end{align}
where ${\rm sig}$ is the signature of a matrix, counting the difference between the number of positive and negative eigenvalues. The numerical computation of the Chern number is made more efficient using LDLT decomposition, see SM~\cite{supp}. The topological index $C_{x,y,E}$ is both space- and energy-resolved and counts the number of boundary modes that cross through $E$; it is $0$ outside of the system and takes a uniform finite value in the bulk, see SM~\cite{supp} for plots of $C_{x,y,E}$ in real space. 
To extract the topological invariant for the whole system, we compute $C_{x,y,E}$ at $E=0,\pi$ for each site within a central region in the bulk and construct their average $C_{0,\pi}$. We further generate many different amorphous configurations and denote the final averages over all configurations as $\bar{C}_{0,\pi}$. We also compute the variance of the Chern numbers $C_{0,\pi}$ in real space, which indicates the stability of the Chern numbers.
Another quantity from the spectral localizer is the localizer gap, which we define in the SM~\cite{supp}.

%~~~~~~~~~~~~~~~~~~~~~~~~~~~~~~~~~~~~~~~~~~~~~~~~~~~~~~~~~
%~~~~~~~~~~~~~~~~~~~~~~~~~~~~~~~~~~~~~~~~~~~~~~~~~~~~~~~~~
%~~~~~~~~~~~~~~~~~~~~~~~~~~~~~~~~~~~~~~~~~~~~~~~~~~~~~~~~~
%~~~~~~~~~~~~~~~~~~~~~~~~~~~~~~~~~~~~~~~~~~~~~~~~~~~~~~~~~

Figure~\ref{Fig:PhaseDiagramAmorphousGraphene} shows $\bar{C}_{0,\pi}$ in the $A \mhyphen \Omega$ plane. First, we focus on crystalline graphene in Figs.~\ref{Fig:PhaseDiagramAmorphousGraphene}(a,b). 
Regions with non-zero $\bar{C}_{0,\pi}$ denote topologically non-trivial graphene regions, generated by the circularly polarized laser.
The phase diagram of irradiated pristine graphene has been studied before~\cite{PerezPRA2015, MikamiBW2016}.
For high frequencies ($\Omega \gg$ bandwidth), we observe non-vanishing $\bar{C}_0=-1$ while $\bar{C}_\pi$ remains zero, indicating that the system only supports topological modes at quasienergy $0$. 
This can be understood as follows. In the high-frequency limit, the Floquet side-bands, which are copies of the bands of the undriven system in frequency space, are far apart and thus overlap or direct transitions between them are forbidden \cite{MikamiBW2016,oka2019,NHLindner2020}. 
However, virtual transitions between different Floquet side-bands may still occur, which can modify the band topology of the system. Thus, in the high-frequency limit, one can obtain an effective Hamiltonian of driven graphene employing perturbation theory, which has been shown to effectively resemble the Haldane model~\cite{MikamiBW2016,oka2019}. 
If we now instead focus on lower to moderate frequencies ($\Omega \simeq$ bandwidth), the phase diagrams of both $\bar{C}_{0}$ and $\bar{C}_{\pi}$ consist of multiple non-trivial regions.
At these lower frequencies, different Floquet side-bands overlap and provide possibilities of band inversion occurring at both quasienergies $0$ and $\pm \pi$~\cite{Kitagawacharacterization2010,JiangColdAtomPRL2011,Rudner2013,Piskunow2014,Usaj2014,PerezPRA2015,Yan2017,Eckardt2017,NHLindner2020,GhoshJPCMReview2024}. 
As a result, the driven system hosts topologically non-trivial edge states at both $0$- and $\pi$-quasienergies. 
For some parameters, we even observe higher Chern numbers, $\left|\bar{C}_{0,\pi}\right| >  1$.

Having discussed the generation of topological phases in crystalline graphene, we now showcase our main result, the phase diagrams for driven monolayer amorphous carbon in Figs.~\ref{Fig:PhaseDiagramAmorphousGraphene}(c,d). Surprisingly, driven monolayer amorphous carbon closely resembles that of crystalline graphene. We only observe a slight reduction in the topological phases. We also shade the phase diagrams (green lines) where the spatial Chern number variance is non-zero, indicating a lack of stability of the corresponding phase. The variance reveals that the phase diagram for $\bar{C}_0$ is very stable apart from around the phase transitions between different Chern numbers. However, for $\bar{C}_\pi$ we find a finite variance in most of the phase diagram, implying that the $\pi$-modes are not very stable, in agreement with the delocalized LDOS of the $\pi$-modes, discussed in Fig.~\ref{Fig:Polycrystalline}(f). Nevertheless, there still exists a small region where $\bar{C}_\pi \neq 0$ with vanishing variance, representing a stable $\pi$-mode, see the SM for variance data~\cite{supp}. Overall, the results demonstrate that monolayer amorphous carbon can easily host the regular $0$-, as well as anomalous $\pi$-modes.

%~~~~~~~~~~~~~~~~~~~~~~~~~~~~~~~~~~~~~~~~~~~~~~~~~~~~~~~~~
%~~~~~~~~~~~~~~~~~~~~~~~~~~~~~~~~~~~~~~~~~~~~~~~~~~~~~~~~~
\begin{figure}
    \centering
    \subfigure{\includegraphics[width=0.48\textwidth]{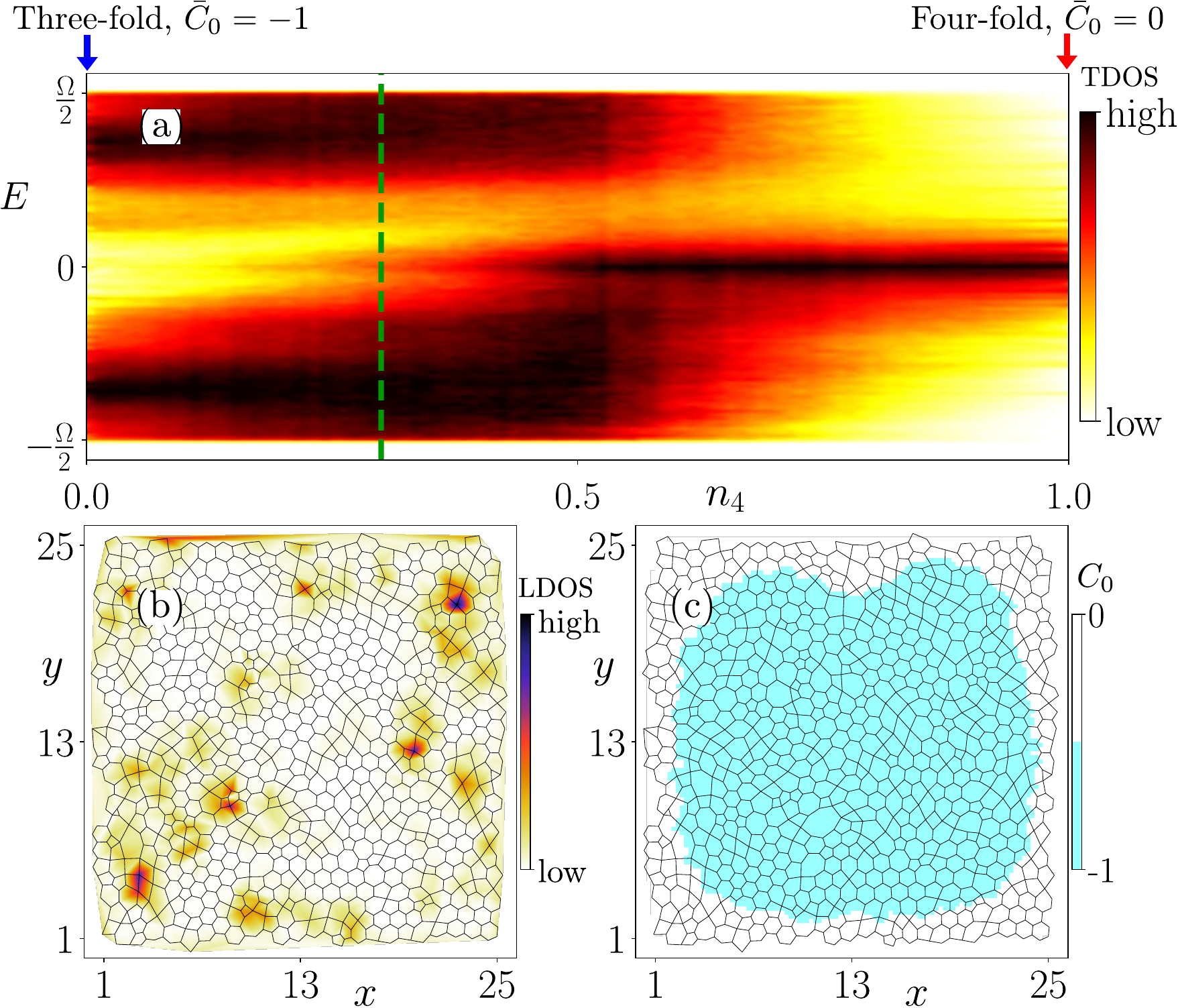}}
    \caption{(a) TDOS as a function of the fraction of fourfold-coordinated sites $n_4$ in driven monolayer amorphous carbon averaged over $50$ random configurations of merging bonds, with all other parameters same as Fig.~\ref{Fig:Polycrystalline}.  Green line indicates the $n_4$ used in (b,c). (b) LDOS and (c) Chern number $C_0$ for the $0$-gap as a function of the system's dimension. For LDOS computation, we use a quasienergy window of $[0.2,0.3]$ in $0$-gap.}
    \label{Fig:TDOSAmorphousBondMerging}
\end{figure}
%~~~~~~~~~~~~~~~~~~~~~~~~~~~~~~~~~~~~~~~~~~~~~~~~~~~~~~~~~
%~~~~~~~~~~~~~~~~~~~~~~~~~~~~~~~~~~~~~~~~~~~~~~~~~~~~~~~~~

\textcolor{blue}{\it Local coordination.---}
So far, we have focused on generating topology in driven monolayer amorphous carbon. 
Although monolayer amorphous carbon consists of polygons different from hexagons, all carbon atoms are still threefold-coordinated.
Next, we establish the importance of the local coordination, here threefold coordination, to generate the non-trivial topology. 
To this end, we consider the addition of fourfold-coordinated defect sites in the threefold-coordinated lattice, experimentally generated by vacancies or entrapment of other atoms~\cite{KrasheninnikovPRL2009, ZhouPRL2012, WangNL2012, RamasseNL2013, YanCSR2018}.
We numerically obtain these defects by merging two neighboring threefold-coordinated sites into one. Further, considering an additional orbital on the fourfold coordinated sites does not give rise to a topological phase. 
Starting from a perfectly threefold-coordinated amorphous lattice, we arbitrarily partition the graph into dimers of neighboring sites and merge a fraction $n_4$ of them chosen at random. 
Thus, $n_4=0$ for a completely threefold-coordinated lattice, as previously, and $n_4=1$ for a fully fourfold-coordinated lattice. 
For intermediate  $n_4$ values, we average over many different configurations of merging bonds.

We start with the same lattice and driving parameters as in Figs.~\ref{Fig:Polycrystalline}(d-e).
We compute the TDOS and observe how it evolves as the fraction of fourfold-coordinated sites $n_4$ is increased in Fig.~\ref{Fig:TDOSAmorphousBondMerging}(a). Since the $\pi$-modes are unstable in this system, we only focus on the topological modes appearing in $0$-gap.
At $n_4=0$, the system has two bands (dark red) separated by a gap which supports gapless edge states, as seen by its finite TDOS (yellow). The system also carries a finite Chern number $\bar{C}_0=-1$ (marked by blue arrow). At low but non-zero $n_4$, the band structure and thus the topology are not affected. 
In this regime, fourfold-coordinated sites just behave as local defects, to which topological properties are robust. 
As the number of fourfold-coordinated sites increases, we observe that the two bands merge into one, around $n_4 \simeq 0.5$. 
This corresponds to the point where the fourfold sites start clustering and cannot be considered as local defects anymore. 
Rather, the lattice is now a mixture of two different local environments, three- and fourfold-coordinated.
As the gap closes, the system also necessarily becomes non-topological, without any edge states. 
Finally, at $n_4=1$, the system consists of fully fourfold-coordinated sites, and it mimics the TDOS of a single band topologically trivial metallic square lattice, with $\bar{C}_0=0$ (marked by red arrow). 
To display the interesting intermediate behavior, we focus on $n_4=0.3$ (green line in Fig.~\ref{Fig:TDOSAmorphousBondMerging}(a)) and show the LDOS computed near the $0$-gap and the Chern number $C_0$ in real space in Figs.~\ref{Fig:TDOSAmorphousBondMerging}(b,c), respectively. The LDOS pattern reveals the existence of edge states, along with some defect states appearing at the fourfold-coordinated sites. The Chern number $C_0$ also gives a clear signature of the non-trivial topology, with $C_0=-1$ inside the system.

Figure~\ref{Fig:TDOSAmorphousBondMerging} discovers a crucial feature of the threefold-coordinated lattice, as in monolayer amorphous carbon. To support a topologically protected edge state, a system must possess at least two bands, which emerge from the sublattice degree of freedom in a threefold-coordinated lattice.
Remarkably, the topological gap survives in the amorphous lattice, despite the two sublattices clearly mixing and thus the bipartition being lost. 
This demonstrates that local coordination is crucial, but that long-range ordering is not necessary to obtain topological states. 
In contrast, fourfold-coordinated lattices host no natural sublattices and thus, only feature one energy band, independent of the system being crystalline or amorphous. Hence, no light-induced topological gap can be opened.

\textcolor{blue}{\textit{Conclusion.---}}
In this work, we utilize a circularly polarized laser to theoretically engineer topological phases in monolayer amorphous carbon, which completely lacks translational symmetry and is topologically trivial in the static limit. In the low-frequency regime, we find both regular $0$- and anomalous $\pi$-modes. We calculate the spectral properties of the driven system with its signature of the edge states and use the spectral localizer to extract an energy- and space-resolved topological invariant to obtain the full phase diagram. Finally, we uncover the crucial importance of local coordination, here the threefold coordination of the monolayer amorphous carbon layer, and establish that the topological phase is destroyed when fourfold-coordinated sites become dominant. 
This reliance on local coordination makes the topology of amorphous materials fundamentally different from that of disordered materials. An interesting future direction is the exploration of driven amorphous systems via signatures such as a quantized orbital magnetization density~\cite{NathanPRL2017,GavenskyPRX2025}.

Experimentally, periodic lasers have already been used to observe topological and Floquet side bands in crystalline graphene~\cite{MerboldtNP2025} and in other materials~\cite{WangScience2013,ZhouNature2023}. Similar experimental setups can also generate and detect the topological edge modes in monolayer amorphous carbon. 
To support this, we estimate the experimental parameters for Fig.~\ref{Fig:PhaseDiagramAmorphousGraphene} in real units~\cite{Bao2022}. Approximating the carbon-carbon hopping amplitude to $t_h=2.7$~eV, the photon energy $E=\hbar\Omega$ is in the range $1.35$ to $27$~eV to reach the topological phase, with the corresponding laser wavelength $\lambda=hc/E$ being $900$~${\rm nm}$ (infrared) to $50$ nm (ultraviolet). The peak electric field intensity $E_0=\hbar \Omega A/e a$ is in the range $10^{9}$ to $10^{11}$~${\rm Vm^{-1}}$, for an estimate of the neighboring carbon-carbon distance $a=1.4$~\AA. The resulting laser field intensity $I=\frac{1}{2} c \epsilon_0 E_0^2$ is then in the range $10^{12}$ to $10^{16}$ ${\rm Wcm^{-2}}$, where $\epsilon_0$ is the vacuum permittivity, resulting in both appropriate laser wavelength and power. At the outset, we note that the damage threshold for crystalline graphene for $800$~${\rm nm}$ light is $3 \times 10^{12}$~${\rm Wcm^{-2}}$~\cite{RobertsAPL2011}. We expect comparable damage thresholds for monolayer amorphous carbon as well. Therefore, intensities in the lower part of the estimated range approaching $10^{13}$~${\rm Wcm^{-2}}$ under short-pulse excitation are likely to be experimentally more viable.
Overall, our work paves the way for controllable topological phases in amorphous systems based on local environment engineering.

\textcolor{black}{\textit{Data Availability.---}}
The data that support the findings of this article are openly available at Ref.~\cite{ghosh_2026_18412848}.

%======================================================
%\subsection*{Acknowledgments}
%======================================================
We acknowledge financial support from the Swedish Research Council (Vetenskapsrådet) Grant No.~2022-03963 and the European Research Council (ERC) under the European Union’s Horizon 2020 research and innovation programme (ERC-2022-CoG, Grant agreement No.~101087096). Part of the computations were enabled by resources provided by the National Academic Infrastructure for Supercomputing in Sweden (NAISS), partially funded by the Swedish Research Council through grant agreement No.~2022-06725. The Python package Kwant~\cite{Groth_2014} has been employed to construct the lattice setups.

\bibliographystyle{apsrev4-2mod}
\bibliography{bibfile.bib}

%apsrev4-2.bst 2015-08-30 from 4.21a (PWD, AO, DPC/HNN) hacked
\begin{thebibliography}{79}%
\makeatletter
\providecommand \@ifxundefined [1]{%
 \@ifx{#1\undefined}
}%
\providecommand \@ifnum [1]{%
 \ifnum #1\expandafter \@firstoftwo
 \else \expandafter \@secondoftwo
 \fi
}%
\providecommand \@ifx [1]{%
 \ifx #1\expandafter \@firstoftwo
 \else \expandafter \@secondoftwo
 \fi
}%
\providecommand \natexlab [1]{#1}%
\providecommand \enquote  [1]{``#1''}%
\providecommand \bibnamefont  [1]{#1}%
\providecommand \bibfnamefont [1]{#1}%
\providecommand \citenamefont [1]{#1}%
\providecommand \href@noop [0]{\@secondoftwo}%
\providecommand \href [0]{\begingroup \@sanitize@url \@href}%
\providecommand \@href[1]{\@@startlink{#1}\@@href}%
\providecommand \@@href[1]{\endgroup#1\@@endlink}%
\providecommand \@sanitize@url [0]{\catcode `\\12\catcode `\$12\catcode
  `\&12\catcode `\#12\catcode `\^12\catcode `\_12\catcode `\%12\relax}%
\providecommand \@@startlink[1]{}%
\providecommand \@@endlink[0]{}%
\providecommand \url  [0]{\begingroup\@sanitize@url \@url }%
\providecommand \@url [1]{\endgroup\@href {#1}{\urlprefix }}%
\providecommand \urlprefix  [0]{URL }%
\providecommand \Eprint [0]{\href }%
\providecommand \doibase [0]{http://dx.doi.org/}%
\providecommand \selectlanguage [0]{\@gobble}%
\providecommand \bibinfo  [0]{\@secondoftwo}%
\providecommand \bibfield  [0]{\@secondoftwo}%
\providecommand \translation [1]{[#1]}%
\providecommand \BibitemOpen [0]{}%
\providecommand \bibitemStop [0]{}%
\providecommand \bibitemNoStop [0]{.\EOS\space}%
\providecommand \EOS [0]{\spacefactor3000\relax}%
\providecommand \BibitemShut  [1]{\csname bibitem#1\endcsname}%
\let\auto@bib@innerbib\@empty
%</preamble>
\bibitem [{\citenamefont {Haldane}(1988)}]{HaldanePRL1988}%
  \BibitemOpen
  \bibfield  {author} {\bibinfo {author} {\bibfnamefont {F.~D.~M.}\
  \bibnamefont {Haldane}},\ }\bibfield  {title} {\emph {\enquote {\bibinfo
  {title} {{Model for a Quantum Hall Effect without Landau Levels:
  Condensed-Matter Realization of the ``Parity Anomaly"}},}\ }}\href {\doibase
  10.1103/PhysRevLett.61.2015} {\bibfield  {journal} {\bibinfo  {journal}
  {Phys. Rev. Lett.}\ }\textbf {\bibinfo {volume} {61}},\ \bibinfo {pages}
  {2015} (\bibinfo {year} {1988})}\BibitemShut {NoStop}%
\bibitem [{\citenamefont {Qi}\ and\ \citenamefont
  {Zhang}(2011)}]{qi2011topological}%
  \BibitemOpen
  \bibfield  {author} {\bibinfo {author} {\bibfnamefont {X.-L.}\ \bibnamefont
  {Qi}}\ and\ \bibinfo {author} {\bibfnamefont {S.-C.}\ \bibnamefont {Zhang}},\
  }\bibfield  {title} {\emph {\enquote {\bibinfo {title} {{Topological
  insulators and superconductors}},}\ }}\href {\doibase
  10.1103/RevModPhys.83.1057} {\bibfield  {journal} {\bibinfo  {journal} {Rev.
  Mod. Phys.}\ }\textbf {\bibinfo {volume} {83}},\ \bibinfo {pages} {1057}
  (\bibinfo {year} {2011})}\BibitemShut {NoStop}%
\bibitem [{\citenamefont {Bernevig}(2013)}]{BernevigBOOK}%
  \BibitemOpen
  \bibfield  {author} {\bibinfo {author} {\bibfnamefont {B.~A.}\ \bibnamefont
  {Bernevig}},\ }\href {\doibase doi:10.1515/9781400846733} {\emph {\bibinfo
  {title} {{Topological Insulators and Topological Superconductors}}}}\
  (\bibinfo  {publisher} {Princeton University Press},\ \bibinfo {address}
  {Princeton},\ \bibinfo {year} {2013})\BibitemShut {NoStop}%
\bibitem [{\citenamefont {Chang}\ \emph {et~al.}(2023)\citenamefont {Chang},
  \citenamefont {Liu},\ and\ \citenamefont {MacDonald}}]{ChangQAHRMP2023}%
  \BibitemOpen
  \bibfield  {author} {\bibinfo {author} {\bibfnamefont {C.-Z.}\ \bibnamefont
  {Chang}}, \bibinfo {author} {\bibfnamefont {C.-X.}\ \bibnamefont {Liu}}, \
  and\ \bibinfo {author} {\bibfnamefont {A.~H.}\ \bibnamefont {MacDonald}},\
  }\bibfield  {title} {\emph {\enquote {\bibinfo {title} {{Colloquium: Quantum
  anomalous Hall effect}},}\ }}\href {\doibase 10.1103/RevModPhys.95.011002}
  {\bibfield  {journal} {\bibinfo  {journal} {Rev. Mod. Phys.}\ }\textbf
  {\bibinfo {volume} {95}},\ \bibinfo {pages} {011002} (\bibinfo {year}
  {2023})}\BibitemShut {NoStop}%
\bibitem [{\citenamefont {Kitagawa}\ \emph {et~al.}(2010)\citenamefont
  {Kitagawa}, \citenamefont {Berg}, \citenamefont {Rudner},\ and\ \citenamefont
  {Demler}}]{Kitagawacharacterization2010}%
  \BibitemOpen
  \bibfield  {author} {\bibinfo {author} {\bibfnamefont {T.}~\bibnamefont
  {Kitagawa}}, \bibinfo {author} {\bibfnamefont {E.}~\bibnamefont {Berg}},
  \bibinfo {author} {\bibfnamefont {M.}~\bibnamefont {Rudner}}, \ and\ \bibinfo
  {author} {\bibfnamefont {E.}~\bibnamefont {Demler}},\ }\bibfield  {title}
  {\emph {\enquote {\bibinfo {title} {Topological characterization of
  periodically driven quantum systems},}\ }}\href {\doibase
  10.1103/PhysRevB.82.235114} {\bibfield  {journal} {\bibinfo  {journal} {Phys.
  Rev. B}\ }\textbf {\bibinfo {volume} {82}},\ \bibinfo {pages} {235114}
  (\bibinfo {year} {2010})}\BibitemShut {NoStop}%
\bibitem [{\citenamefont {Kitagawa}\ \emph {et~al.}(2011)\citenamefont
  {Kitagawa}, \citenamefont {Oka}, \citenamefont {Brataas}, \citenamefont
  {Fu},\ and\ \citenamefont {Demler}}]{kitagawa11transport}%
  \BibitemOpen
  \bibfield  {author} {\bibinfo {author} {\bibfnamefont {T.}~\bibnamefont
  {Kitagawa}}, \bibinfo {author} {\bibfnamefont {T.}~\bibnamefont {Oka}},
  \bibinfo {author} {\bibfnamefont {A.}~\bibnamefont {Brataas}}, \bibinfo
  {author} {\bibfnamefont {L.}~\bibnamefont {Fu}}, \ and\ \bibinfo {author}
  {\bibfnamefont {E.}~\bibnamefont {Demler}},\ }\bibfield  {title} {\emph
  {\enquote {\bibinfo {title} {{Transport properties of nonequilibrium systems
  under the application of light: Photoinduced quantum Hall insulators without
  Landau levels}},}\ }}\href {\doibase 10.1103/PhysRevB.84.235108} {\bibfield
  {journal} {\bibinfo  {journal} {Phys. Rev. B}\ }\textbf {\bibinfo {volume}
  {84}},\ \bibinfo {pages} {235108} (\bibinfo {year} {2011})}\BibitemShut
  {NoStop}%
\bibitem [{\citenamefont {Lindner}\ \emph {et~al.}(2011)\citenamefont
  {Lindner}, \citenamefont {Refael},\ and\ \citenamefont
  {Galitski}}]{lindner11floquet}%
  \BibitemOpen
  \bibfield  {author} {\bibinfo {author} {\bibfnamefont {N.~H.}\ \bibnamefont
  {Lindner}}, \bibinfo {author} {\bibfnamefont {G.}~\bibnamefont {Refael}}, \
  and\ \bibinfo {author} {\bibfnamefont {V.}~\bibnamefont {Galitski}},\
  }\bibfield  {title} {\emph {\enquote {\bibinfo {title} {Floquet topological
  insulator in semiconductor quantum wells},}\ }}\href {\doibase
  https://doi.org/10.1038/nphys1926} {\bibfield  {journal} {\bibinfo  {journal}
  {Nature Phys}\ }\textbf {\bibinfo {volume} {7}},\ \bibinfo {pages} {490}
  (\bibinfo {year} {2011})}\BibitemShut {NoStop}%
\bibitem [{\citenamefont {Gu}\ \emph {et~al.}(2011)\citenamefont {Gu},
  \citenamefont {Fertig}, \citenamefont {Arovas},\ and\ \citenamefont
  {Auerbach}}]{FloquetGuPRL2011}%
  \BibitemOpen
  \bibfield  {author} {\bibinfo {author} {\bibfnamefont {Z.}~\bibnamefont
  {Gu}}, \bibinfo {author} {\bibfnamefont {H.~A.}\ \bibnamefont {Fertig}},
  \bibinfo {author} {\bibfnamefont {D.~P.}\ \bibnamefont {Arovas}}, \ and\
  \bibinfo {author} {\bibfnamefont {A.}~\bibnamefont {Auerbach}},\ }\bibfield
  {title} {\emph {\enquote {\bibinfo {title} {Floquet Spectrum and Transport
  through an Irradiated Graphene Ribbon},}\ }}\href {\doibase
  10.1103/PhysRevLett.107.216601} {\bibfield  {journal} {\bibinfo  {journal}
  {Phys. Rev. Lett.}\ }\textbf {\bibinfo {volume} {107}},\ \bibinfo {pages}
  {216601} (\bibinfo {year} {2011})}\BibitemShut {NoStop}%
\bibitem [{\citenamefont {Rudner}\ \emph {et~al.}(2013)\citenamefont {Rudner},
  \citenamefont {Lindner}, \citenamefont {Berg},\ and\ \citenamefont
  {Levin}}]{Rudner2013}%
  \BibitemOpen
  \bibfield  {author} {\bibinfo {author} {\bibfnamefont {M.~S.}\ \bibnamefont
  {Rudner}}, \bibinfo {author} {\bibfnamefont {N.~H.}\ \bibnamefont {Lindner}},
  \bibinfo {author} {\bibfnamefont {E.}~\bibnamefont {Berg}}, \ and\ \bibinfo
  {author} {\bibfnamefont {M.}~\bibnamefont {Levin}},\ }\bibfield  {title}
  {\emph {\enquote {\bibinfo {title} {{Anomalous Edge States and the Bulk-Edge
  Correspondence for Periodically Driven Two-Dimensional Systems}},}\ }}\href
  {\doibase 10.1103/PhysRevX.3.031005} {\bibfield  {journal} {\bibinfo
  {journal} {Phys. Rev. X}\ }\textbf {\bibinfo {volume} {3}},\ \bibinfo {pages}
  {031005} (\bibinfo {year} {2013})}\BibitemShut {NoStop}%
\bibitem [{\citenamefont {Usaj}\ \emph {et~al.}(2014)\citenamefont {Usaj},
  \citenamefont {Perez-Piskunow}, \citenamefont {Foa~Torres},\ and\
  \citenamefont {Balseiro}}]{Usaj2014}%
  \BibitemOpen
  \bibfield  {author} {\bibinfo {author} {\bibfnamefont {G.}~\bibnamefont
  {Usaj}}, \bibinfo {author} {\bibfnamefont {P.~M.}\ \bibnamefont
  {Perez-Piskunow}}, \bibinfo {author} {\bibfnamefont {L.~E.~F.}\ \bibnamefont
  {Foa~Torres}}, \ and\ \bibinfo {author} {\bibfnamefont {C.~A.}\ \bibnamefont
  {Balseiro}},\ }\bibfield  {title} {\emph {\enquote {\bibinfo {title}
  {{Irradiated graphene as a tunable Floquet topological insulator}},}\ }}\href
  {\doibase 10.1103/PhysRevB.90.115423} {\bibfield  {journal} {\bibinfo
  {journal} {Phys. Rev. B}\ }\textbf {\bibinfo {volume} {90}},\ \bibinfo
  {pages} {115423} (\bibinfo {year} {2014})}\BibitemShut {NoStop}%
\bibitem [{\citenamefont {Perez-Piskunow}\ \emph {et~al.}(2015)\citenamefont
  {Perez-Piskunow}, \citenamefont {Foa~Torres},\ and\ \citenamefont
  {Usaj}}]{PerezPRA2015}%
  \BibitemOpen
  \bibfield  {author} {\bibinfo {author} {\bibfnamefont {P.~M.}\ \bibnamefont
  {Perez-Piskunow}}, \bibinfo {author} {\bibfnamefont {L.~E.~F.}\ \bibnamefont
  {Foa~Torres}}, \ and\ \bibinfo {author} {\bibfnamefont {G.}~\bibnamefont
  {Usaj}},\ }\bibfield  {title} {\emph {\enquote {\bibinfo {title} {Hierarchy
  of Floquet gaps and edge states for driven honeycomb lattices},}\ }}\href
  {\doibase 10.1103/PhysRevA.91.043625} {\bibfield  {journal} {\bibinfo
  {journal} {Phys. Rev. A}\ }\textbf {\bibinfo {volume} {91}},\ \bibinfo
  {pages} {043625} (\bibinfo {year} {2015})}\BibitemShut {NoStop}%
\bibitem [{\citenamefont {Eckardt}(2017)}]{Eckardt2017}%
  \BibitemOpen
  \bibfield  {author} {\bibinfo {author} {\bibfnamefont {A.}~\bibnamefont
  {Eckardt}},\ }\bibfield  {title} {\emph {\enquote {\bibinfo {title}
  {{Colloquium: Atomic quantum gases in periodically driven optical
  lattices}},}\ }}\href {\doibase 10.1103/RevModPhys.89.011004} {\bibfield
  {journal} {\bibinfo  {journal} {Rev. Mod. Phys.}\ }\textbf {\bibinfo {volume}
  {89}},\ \bibinfo {pages} {011004} (\bibinfo {year} {2017})}\BibitemShut
  {NoStop}%
\bibitem [{\citenamefont {Oka}\ and\ \citenamefont {Kitamura}(2019)}]{oka2019}%
  \BibitemOpen
  \bibfield  {author} {\bibinfo {author} {\bibfnamefont {T.}~\bibnamefont
  {Oka}}\ and\ \bibinfo {author} {\bibfnamefont {S.}~\bibnamefont {Kitamura}},\
  }\bibfield  {title} {\emph {\enquote {\bibinfo {title} {{Floquet Engineering
  of Quantum Materials}},}\ }}\href {\doibase
  10.1146/annurev-conmatphys-031218-013423} {\bibfield  {journal} {\bibinfo
  {journal} {Annu. Rev. Condens. Matter Phys.}\ }\textbf {\bibinfo {volume}
  {10}},\ \bibinfo {pages} {387} (\bibinfo {year} {2019})}\BibitemShut
  {NoStop}%
\bibitem [{\citenamefont {Rudner}\ and\ \citenamefont
  {Lindner}(2020)}]{NHLindner2020}%
  \BibitemOpen
  \bibfield  {author} {\bibinfo {author} {\bibfnamefont {M.}~\bibnamefont
  {Rudner}}\ and\ \bibinfo {author} {\bibfnamefont {N.}~\bibnamefont
  {Lindner}},\ }\bibfield  {title} {\emph {\enquote {\bibinfo {title} {{Band
  structure engineering and non-equilibrium dynamics in Floquet topological
  insulators}},}\ }}\href {\doibase https://doi.org/10.1038/s42254-020-0170-z}
  {\bibfield  {journal} {\bibinfo  {journal} {Nat. Rev. Phys.}\ }\textbf
  {\bibinfo {volume} {2}},\ \bibinfo {pages} {229–244} (\bibinfo {year}
  {2020})}\BibitemShut {NoStop}%
\bibitem [{\citenamefont {Bao}\ \emph {et~al.}(2022)\citenamefont {Bao},
  \citenamefont {Tang}, \citenamefont {Sun},\ and\ \citenamefont
  {Zhou}}]{Bao2022}%
  \BibitemOpen
  \bibfield  {author} {\bibinfo {author} {\bibfnamefont {C.}~\bibnamefont
  {Bao}}, \bibinfo {author} {\bibfnamefont {P.}~\bibnamefont {Tang}}, \bibinfo
  {author} {\bibfnamefont {D.}~\bibnamefont {Sun}}, \ and\ \bibinfo {author}
  {\bibfnamefont {S.}~\bibnamefont {Zhou}},\ }\bibfield  {title} {\emph
  {\enquote {\bibinfo {title} {Light-induced emergent phenomena in 2D materials
  and topological materials},}\ }}\href {\doibase 10.1038/s42254-021-00388-1}
  {\bibfield  {journal} {\bibinfo  {journal} {Nat. Rev. Phys.}\ }\textbf
  {\bibinfo {volume} {4}},\ \bibinfo {pages} {33} (\bibinfo {year}
  {2022})}\BibitemShut {NoStop}%
\bibitem [{\citenamefont {Ghosh}\ \emph {et~al.}(2023)\citenamefont {Ghosh},
  \citenamefont {Nag},\ and\ \citenamefont {Saha}}]{GhoshJPCMReview2024}%
  \BibitemOpen
  \bibfield  {author} {\bibinfo {author} {\bibfnamefont {A.~K.}\ \bibnamefont
  {Ghosh}}, \bibinfo {author} {\bibfnamefont {T.}~\bibnamefont {Nag}}, \ and\
  \bibinfo {author} {\bibfnamefont {A.}~\bibnamefont {Saha}},\ }\bibfield
  {title} {\emph {\enquote {\bibinfo {title} {Generation of higher-order
  topological insulators using periodic driving},}\ }}\href {\doibase
  10.1088/1361-648X/ad0e2d} {\bibfield  {journal} {\bibinfo  {journal} {J.
  Phys.: Condens. Matter}\ }\textbf {\bibinfo {volume} {36}},\ \bibinfo {pages}
  {093001} (\bibinfo {year} {2023})}\BibitemShut {NoStop}%
\bibitem [{\citenamefont {Jiang}\ \emph {et~al.}(2011)\citenamefont {Jiang},
  \citenamefont {Kitagawa}, \citenamefont {Alicea}, \citenamefont {Akhmerov},
  \citenamefont {Pekker}, \citenamefont {Refael}, \citenamefont {Cirac},
  \citenamefont {Demler}, \citenamefont {Lukin},\ and\ \citenamefont
  {Zoller}}]{JiangColdAtomPRL2011}%
  \BibitemOpen
  \bibfield  {author} {\bibinfo {author} {\bibfnamefont {L.}~\bibnamefont
  {Jiang}}, \bibinfo {author} {\bibfnamefont {T.}~\bibnamefont {Kitagawa}},
  \bibinfo {author} {\bibfnamefont {J.}~\bibnamefont {Alicea}}, \bibinfo
  {author} {\bibfnamefont {A.~R.}\ \bibnamefont {Akhmerov}}, \bibinfo {author}
  {\bibfnamefont {D.}~\bibnamefont {Pekker}}, \bibinfo {author} {\bibfnamefont
  {G.}~\bibnamefont {Refael}}, \bibinfo {author} {\bibfnamefont {J.~I.}\
  \bibnamefont {Cirac}}, \bibinfo {author} {\bibfnamefont {E.}~\bibnamefont
  {Demler}}, \bibinfo {author} {\bibfnamefont {M.~D.}\ \bibnamefont {Lukin}}, \
  and\ \bibinfo {author} {\bibfnamefont {P.}~\bibnamefont {Zoller}},\
  }\bibfield  {title} {\emph {\enquote {\bibinfo {title} {Majorana Fermions in
  Equilibrium and in Driven Cold-Atom Quantum Wires},}\ }}\href {\doibase
  10.1103/PhysRevLett.106.220402} {\bibfield  {journal} {\bibinfo  {journal}
  {Phys. Rev. Lett.}\ }\textbf {\bibinfo {volume} {106}},\ \bibinfo {pages}
  {220402} (\bibinfo {year} {2011})}\BibitemShut {NoStop}%
\bibitem [{\citenamefont {Perez-Piskunow}\ \emph {et~al.}(2014)\citenamefont
  {Perez-Piskunow}, \citenamefont {Usaj}, \citenamefont {Balseiro},\ and\
  \citenamefont {Torres}}]{Piskunow2014}%
  \BibitemOpen
  \bibfield  {author} {\bibinfo {author} {\bibfnamefont {P.~M.}\ \bibnamefont
  {Perez-Piskunow}}, \bibinfo {author} {\bibfnamefont {G.}~\bibnamefont
  {Usaj}}, \bibinfo {author} {\bibfnamefont {C.~A.}\ \bibnamefont {Balseiro}},
  \ and\ \bibinfo {author} {\bibfnamefont {L.~E. F.~F.}\ \bibnamefont
  {Torres}},\ }\bibfield  {title} {\emph {\enquote {\bibinfo {title} {Floquet
  chiral edge states in graphene},}\ }}\href {\doibase
  10.1103/PhysRevB.89.121401} {\bibfield  {journal} {\bibinfo  {journal} {Phys.
  Rev. B}\ }\textbf {\bibinfo {volume} {89}},\ \bibinfo {pages} {121401}
  (\bibinfo {year} {2014})}\BibitemShut {NoStop}%
\bibitem [{\citenamefont {Yao}\ \emph {et~al.}(2017)\citenamefont {Yao},
  \citenamefont {Yan},\ and\ \citenamefont {Wang}}]{Yan2017}%
  \BibitemOpen
  \bibfield  {author} {\bibinfo {author} {\bibfnamefont {S.}~\bibnamefont
  {Yao}}, \bibinfo {author} {\bibfnamefont {Z.}~\bibnamefont {Yan}}, \ and\
  \bibinfo {author} {\bibfnamefont {Z.}~\bibnamefont {Wang}},\ }\bibfield
  {title} {\emph {\enquote {\bibinfo {title} {{Topological invariants of
  Floquet systems: General formulation, special properties, and Floquet
  topological defects}},}\ }}\href {\doibase 10.1103/PhysRevB.96.195303}
  {\bibfield  {journal} {\bibinfo  {journal} {Phys. Rev. B}\ }\textbf {\bibinfo
  {volume} {96}},\ \bibinfo {pages} {195303} (\bibinfo {year}
  {2017})}\BibitemShut {NoStop}%
\bibitem [{\citenamefont {McIver}\ \emph {et~al.}(2020)\citenamefont {McIver},
  \citenamefont {Schulte}, \citenamefont {Stein}, \citenamefont {Matsuyama},
  \citenamefont {Jotzu}, \citenamefont {Meier},\ and\ \citenamefont
  {Cavalleri}}]{McIver2020}%
  \BibitemOpen
  \bibfield  {author} {\bibinfo {author} {\bibfnamefont {J.~W.}\ \bibnamefont
  {McIver}}, \bibinfo {author} {\bibfnamefont {B.}~\bibnamefont {Schulte}},
  \bibinfo {author} {\bibfnamefont {F.-U.}\ \bibnamefont {Stein}}, \bibinfo
  {author} {\bibfnamefont {T.}~\bibnamefont {Matsuyama}}, \bibinfo {author}
  {\bibfnamefont {G.}~\bibnamefont {Jotzu}}, \bibinfo {author} {\bibfnamefont
  {G.}~\bibnamefont {Meier}}, \ and\ \bibinfo {author} {\bibfnamefont
  {A.}~\bibnamefont {Cavalleri}},\ }\bibfield  {title} {\emph {\enquote
  {\bibinfo {title} {Light-induced anomalous Hall effect in graphene},}\
  }}\href {\doibase 10.1038/s41567-019-0698-y} {\bibfield  {journal} {\bibinfo
  {journal} {Nat. Phys.}\ }\textbf {\bibinfo {volume} {16}},\ \bibinfo {pages}
  {38} (\bibinfo {year} {2020})}\BibitemShut {NoStop}%
\bibitem [{\citenamefont {Lesko}\ \emph {et~al.}()\citenamefont {Lesko},
  \citenamefont {Weitz}, \citenamefont {Wittigschlager}, \citenamefont {Li},
  \citenamefont {Heide}, \citenamefont {Neufeld},\ and\ \citenamefont
  {Hommelhoff}}]{lesko2025opticalcontrolelectronsfloquet}%
  \BibitemOpen
  \bibfield  {author} {\bibinfo {author} {\bibfnamefont {D.~M.~B.}\
  \bibnamefont {Lesko}}, \bibinfo {author} {\bibfnamefont {T.}~\bibnamefont
  {Weitz}}, \bibinfo {author} {\bibfnamefont {S.}~\bibnamefont
  {Wittigschlager}}, \bibinfo {author} {\bibfnamefont {W.}~\bibnamefont {Li}},
  \bibinfo {author} {\bibfnamefont {C.}~\bibnamefont {Heide}}, \bibinfo
  {author} {\bibfnamefont {O.}~\bibnamefont {Neufeld}}, \ and\ \bibinfo
  {author} {\bibfnamefont {P.}~\bibnamefont {Hommelhoff}},\ }\href@noop {}
  {\enquote {\bibinfo {title} {Optical control of electrons in a Floquet
  topological insulator},}\ }\Eprint
  {http://arxiv.org/abs/2407.17917}{arXiv:2407.17917}\BibitemShut {NoStop}%
\bibitem [{\citenamefont {Rechtsman}\ \emph {et~al.}(2013)\citenamefont
  {Rechtsman}, \citenamefont {Zeuner}, \citenamefont {Plotnik}, \citenamefont
  {Lumer}, \citenamefont {Podolsky}, \citenamefont {Dreisow}, \citenamefont
  {Nolte}, \citenamefont {Segev},\ and\ \citenamefont
  {Szameit}}]{RechtsmanExperiment2013}%
  \BibitemOpen
  \bibfield  {author} {\bibinfo {author} {\bibfnamefont {M.~C.}\ \bibnamefont
  {Rechtsman}}, \bibinfo {author} {\bibfnamefont {J.~M.}\ \bibnamefont
  {Zeuner}}, \bibinfo {author} {\bibfnamefont {Y.}~\bibnamefont {Plotnik}},
  \bibinfo {author} {\bibfnamefont {Y.}~\bibnamefont {Lumer}}, \bibinfo
  {author} {\bibfnamefont {D.}~\bibnamefont {Podolsky}}, \bibinfo {author}
  {\bibfnamefont {F.}~\bibnamefont {Dreisow}}, \bibinfo {author} {\bibfnamefont
  {S.}~\bibnamefont {Nolte}}, \bibinfo {author} {\bibfnamefont
  {M.}~\bibnamefont {Segev}}, \ and\ \bibinfo {author} {\bibfnamefont
  {A.}~\bibnamefont {Szameit}},\ }\bibfield  {title} {\emph {\enquote {\bibinfo
  {title} {{Photonic Floquet topological insulators}},}\ }}\href {\doibase
  10.1038/nature12066} {\bibfield  {journal} {\bibinfo  {journal} {Nature}\
  }\textbf {\bibinfo {volume} {496}},\ \bibinfo {pages} {196} (\bibinfo {year}
  {2013})}\BibitemShut {NoStop}%
\bibitem [{\citenamefont {Maczewsky}\ \emph {et~al.}(2017)\citenamefont
  {Maczewsky}, \citenamefont {Zeuner}, \citenamefont {Nolte},\ and\
  \citenamefont {Szameit}}]{Maczewsky2017}%
  \BibitemOpen
  \bibfield  {author} {\bibinfo {author} {\bibfnamefont {L.~J.}\ \bibnamefont
  {Maczewsky}}, \bibinfo {author} {\bibfnamefont {J.~M.}\ \bibnamefont
  {Zeuner}}, \bibinfo {author} {\bibfnamefont {S.}~\bibnamefont {Nolte}}, \
  and\ \bibinfo {author} {\bibfnamefont {A.}~\bibnamefont {Szameit}},\
  }\bibfield  {title} {\emph {\enquote {\bibinfo {title} {Observation of
  photonic anomalous Floquet topological insulators},}\ }}\href {\doibase
  10.1038/ncomms13756} {\bibfield  {journal} {\bibinfo  {journal} {Nat Commun}\
  }\textbf {\bibinfo {volume} {8}},\ \bibinfo {pages} {13756} (\bibinfo {year}
  {2017})}\BibitemShut {NoStop}%
\bibitem [{\citenamefont {Peng}\ \emph {et~al.}(2016)\citenamefont {Peng},
  \citenamefont {Qin}, \citenamefont {Zhao}, \citenamefont {Shen},
  \citenamefont {Xu}, \citenamefont {Bao}, \citenamefont {Jia},\ and\
  \citenamefont {Zhu}}]{Peng2016}%
  \BibitemOpen
  \bibfield  {author} {\bibinfo {author} {\bibfnamefont {Y.-G.}\ \bibnamefont
  {Peng}}, \bibinfo {author} {\bibfnamefont {C.-Z.}\ \bibnamefont {Qin}},
  \bibinfo {author} {\bibfnamefont {D.-G.}\ \bibnamefont {Zhao}}, \bibinfo
  {author} {\bibfnamefont {Y.-X.}\ \bibnamefont {Shen}}, \bibinfo {author}
  {\bibfnamefont {X.-Y.}\ \bibnamefont {Xu}}, \bibinfo {author} {\bibfnamefont
  {M.}~\bibnamefont {Bao}}, \bibinfo {author} {\bibfnamefont {H.}~\bibnamefont
  {Jia}}, \ and\ \bibinfo {author} {\bibfnamefont {X.-F.}\ \bibnamefont
  {Zhu}},\ }\bibfield  {title} {\emph {\enquote {\bibinfo {title} {Experimental
  demonstration of anomalous Floquet topological insulator for sound},}\
  }}\href {\doibase 10.1038/ncomms13368} {\bibfield  {journal} {\bibinfo
  {journal} {Nat Commun}\ }\textbf {\bibinfo {volume} {7}},\ \bibinfo {pages}
  {13368} (\bibinfo {year} {2016})}\BibitemShut {NoStop}%
\bibitem [{\citenamefont {Fleury}\ \emph {et~al.}(2016)\citenamefont {Fleury},
  \citenamefont {Khanikaev},\ and\ \citenamefont {Alu}}]{fleury2016floquet}%
  \BibitemOpen
  \bibfield  {author} {\bibinfo {author} {\bibfnamefont {R.}~\bibnamefont
  {Fleury}}, \bibinfo {author} {\bibfnamefont {A.~B.}\ \bibnamefont
  {Khanikaev}}, \ and\ \bibinfo {author} {\bibfnamefont {A.}~\bibnamefont
  {Alu}},\ }\bibfield  {title} {\emph {\enquote {\bibinfo {title} {Floquet
  topological insulators for sound},}\ }}\href {\doibase 10.1038/ncomms11744}
  {\bibfield  {journal} {\bibinfo  {journal} {Nat Commun}\ }\textbf {\bibinfo
  {volume} {7}},\ \bibinfo {pages} {11744} (\bibinfo {year}
  {2016})}\BibitemShut {NoStop}%
\bibitem [{\citenamefont {Jotzu}\ \emph {et~al.}(2014)\citenamefont {Jotzu},
  \citenamefont {Messer}, \citenamefont {Desbuquois}, \citenamefont {Lebrat},
  \citenamefont {Uehlinger}, \citenamefont {Greif},\ and\ \citenamefont
  {Esslinger}}]{Jotzu2014}%
  \BibitemOpen
  \bibfield  {author} {\bibinfo {author} {\bibfnamefont {G.}~\bibnamefont
  {Jotzu}}, \bibinfo {author} {\bibfnamefont {M.}~\bibnamefont {Messer}},
  \bibinfo {author} {\bibfnamefont {R.}~\bibnamefont {Desbuquois}}, \bibinfo
  {author} {\bibfnamefont {M.}~\bibnamefont {Lebrat}}, \bibinfo {author}
  {\bibfnamefont {T.}~\bibnamefont {Uehlinger}}, \bibinfo {author}
  {\bibfnamefont {D.}~\bibnamefont {Greif}}, \ and\ \bibinfo {author}
  {\bibfnamefont {T.}~\bibnamefont {Esslinger}},\ }\bibfield  {title} {\emph
  {\enquote {\bibinfo {title} {Experimental realization of the topological
  Haldane model with ultracold fermions},}\ }}\href {\doibase
  10.1038/nature13915} {\bibfield  {journal} {\bibinfo  {journal} {Nature}\
  }\textbf {\bibinfo {volume} {515}},\ \bibinfo {pages} {237} (\bibinfo {year}
  {2014})}\BibitemShut {NoStop}%
\bibitem [{\citenamefont {Wintersperger}\ \emph {et~al.}(2020)\citenamefont
  {Wintersperger}, \citenamefont {Braun}, \citenamefont {{\"U}nal},
  \citenamefont {Eckardt}, \citenamefont {Liberto}, \citenamefont {Goldman},
  \citenamefont {Bloch},\ and\ \citenamefont
  {Aidelsburger}}]{Wintersperger2020}%
  \BibitemOpen
  \bibfield  {author} {\bibinfo {author} {\bibfnamefont {K.}~\bibnamefont
  {Wintersperger}}, \bibinfo {author} {\bibfnamefont {C.}~\bibnamefont
  {Braun}}, \bibinfo {author} {\bibfnamefont {F.~N.}\ \bibnamefont {{\"U}nal}},
  \bibinfo {author} {\bibfnamefont {A.}~\bibnamefont {Eckardt}}, \bibinfo
  {author} {\bibfnamefont {M.~D.}\ \bibnamefont {Liberto}}, \bibinfo {author}
  {\bibfnamefont {N.}~\bibnamefont {Goldman}}, \bibinfo {author} {\bibfnamefont
  {I.}~\bibnamefont {Bloch}}, \ and\ \bibinfo {author} {\bibfnamefont
  {M.}~\bibnamefont {Aidelsburger}},\ }\bibfield  {title} {\emph {\enquote
  {\bibinfo {title} {Realization of an anomalous Floquet topological system
  with ultracold atoms},}\ }}\href {\doibase 10.1038/s41567-020-0949-y}
  {\bibfield  {journal} {\bibinfo  {journal} {Nat. Phys.}\ }\textbf {\bibinfo
  {volume} {16}},\ \bibinfo {pages} {1058} (\bibinfo {year}
  {2020})}\BibitemShut {NoStop}%
\bibitem [{\citenamefont {Fu}\ and\ \citenamefont {Kane}(2007)}]{FuKane2007}%
  \BibitemOpen
  \bibfield  {author} {\bibinfo {author} {\bibfnamefont {L.}~\bibnamefont
  {Fu}}\ and\ \bibinfo {author} {\bibfnamefont {C.~L.}\ \bibnamefont {Kane}},\
  }\bibfield  {title} {\emph {\enquote {\bibinfo {title} {{Topological
  insulators with inversion symmetry}},}\ }}\href {\doibase
  10.1103/PhysRevB.76.045302} {\bibfield  {journal} {\bibinfo  {journal} {Phys.
  Rev. B}\ }\textbf {\bibinfo {volume} {76}},\ \bibinfo {pages} {045302}
  (\bibinfo {year} {2007})}\BibitemShut {NoStop}%
\bibitem [{\citenamefont {Po}\ \emph {et~al.}(2017)\citenamefont {Po},
  \citenamefont {Vishwanath},\ and\ \citenamefont
  {Watanabe}}]{PoNatCommun2017}%
  \BibitemOpen
  \bibfield  {author} {\bibinfo {author} {\bibfnamefont {H.~C.}\ \bibnamefont
  {Po}}, \bibinfo {author} {\bibfnamefont {A.}~\bibnamefont {Vishwanath}}, \
  and\ \bibinfo {author} {\bibfnamefont {H.}~\bibnamefont {Watanabe}},\
  }\bibfield  {title} {\emph {\enquote {\bibinfo {title} {Symmetry-based
  indicators of band topology in the 230 space groups},}\ }}\href {\doibase
  10.1038/s41467-017-00133-2} {\bibfield  {journal} {\bibinfo  {journal} {Nat
  Commun}\ }\textbf {\bibinfo {volume} {8}},\ \bibinfo {pages} {50} (\bibinfo
  {year} {2017})}\BibitemShut {NoStop}%
\bibitem [{\citenamefont {Kruthoff}\ \emph {et~al.}(2017)\citenamefont
  {Kruthoff}, \citenamefont {de~Boer}, \citenamefont {van Wezel}, \citenamefont
  {Kane},\ and\ \citenamefont {Slager}}]{SlagerPRX2017}%
  \BibitemOpen
  \bibfield  {author} {\bibinfo {author} {\bibfnamefont {J.}~\bibnamefont
  {Kruthoff}}, \bibinfo {author} {\bibfnamefont {J.}~\bibnamefont {de~Boer}},
  \bibinfo {author} {\bibfnamefont {J.}~\bibnamefont {van Wezel}}, \bibinfo
  {author} {\bibfnamefont {C.~L.}\ \bibnamefont {Kane}}, \ and\ \bibinfo
  {author} {\bibfnamefont {R.-J.}\ \bibnamefont {Slager}},\ }\bibfield  {title}
  {\emph {\enquote {\bibinfo {title} {Topological Classification of Crystalline
  Insulators through Band Structure Combinatorics},}\ }}\href {\doibase
  10.1103/PhysRevX.7.041069} {\bibfield  {journal} {\bibinfo  {journal} {Phys.
  Rev. X}\ }\textbf {\bibinfo {volume} {7}},\ \bibinfo {pages} {041069}
  (\bibinfo {year} {2017})}\BibitemShut {NoStop}%
\bibitem [{\citenamefont {Bradlyn}\ \emph {et~al.}(2017)\citenamefont
  {Bradlyn}, \citenamefont {Elcoro}, \citenamefont {Cano}, \citenamefont
  {Vergniory}, \citenamefont {Wang}, \citenamefont {Felser}, \citenamefont
  {Aroyo},\ and\ \citenamefont {Bernevig}}]{bradlyn2017}%
  \BibitemOpen
  \bibfield  {author} {\bibinfo {author} {\bibfnamefont {B.}~\bibnamefont
  {Bradlyn}}, \bibinfo {author} {\bibfnamefont {L.}~\bibnamefont {Elcoro}},
  \bibinfo {author} {\bibfnamefont {J.}~\bibnamefont {Cano}}, \bibinfo {author}
  {\bibfnamefont {M.~G.}\ \bibnamefont {Vergniory}}, \bibinfo {author}
  {\bibfnamefont {Z.}~\bibnamefont {Wang}}, \bibinfo {author} {\bibfnamefont
  {C.}~\bibnamefont {Felser}}, \bibinfo {author} {\bibfnamefont {M.~I.}\
  \bibnamefont {Aroyo}}, \ and\ \bibinfo {author} {\bibfnamefont {B.~A.}\
  \bibnamefont {Bernevig}},\ }\bibfield  {title} {\emph {\enquote {\bibinfo
  {title} {Topological quantum chemistry},}\ }}\href {\doibase
  https://doi.org/10.1038/nature23268} {\bibfield  {journal} {\bibinfo
  {journal} {Nature}\ }\textbf {\bibinfo {volume} {547}},\ \bibinfo {pages}
  {298} (\bibinfo {year} {2017})}\BibitemShut {NoStop}%
\bibitem [{\citenamefont {Corbae}\ \emph
  {et~al.}(2023{\natexlab{a}})\citenamefont {Corbae}, \citenamefont
  {Hannukainen}, \citenamefont {Marsal}, \citenamefont {Muñoz-Segovia},\ and\
  \citenamefont {Grushin}}]{CorbaeEPL2023}%
  \BibitemOpen
  \bibfield  {author} {\bibinfo {author} {\bibfnamefont {P.}~\bibnamefont
  {Corbae}}, \bibinfo {author} {\bibfnamefont {J.~D.}\ \bibnamefont
  {Hannukainen}}, \bibinfo {author} {\bibfnamefont {Q.}~\bibnamefont {Marsal}},
  \bibinfo {author} {\bibfnamefont {D.}~\bibnamefont {Muñoz-Segovia}}, \ and\
  \bibinfo {author} {\bibfnamefont {A.~G.}\ \bibnamefont {Grushin}},\
  }\bibfield  {title} {\emph {\enquote {\bibinfo {title} {Amorphous topological
  matter: Theory and experiment},}\ }}\href {\doibase 10.1209/0295-5075/acc2e2}
  {\bibfield  {journal} {\bibinfo  {journal} {EPL}\ }\textbf {\bibinfo {volume}
  {142}},\ \bibinfo {pages} {16001} (\bibinfo {year}
  {2023}{\natexlab{a}})}\BibitemShut {NoStop}%
\bibitem [{\citenamefont {Zallen}(1998)}]{zallen_physics_1998}%
  \BibitemOpen
  \bibfield  {author} {\bibinfo {author} {\bibfnamefont {R.}~\bibnamefont
  {Zallen}},\ }\href {\doibase https://doi.org/10.1002/9783527617968} {\emph
  {\bibinfo {title} {The Physics of Amorphous Solids}}}\ (\bibinfo  {publisher}
  {John Wiley \& Sons, Ltd},\ \bibinfo {year} {1998})\BibitemShut {NoStop}%
\bibitem [{\citenamefont {Agarwala}\ and\ \citenamefont
  {Shenoy}(2017)}]{AgarwalaPRL2017}%
  \BibitemOpen
  \bibfield  {author} {\bibinfo {author} {\bibfnamefont {A.}~\bibnamefont
  {Agarwala}}\ and\ \bibinfo {author} {\bibfnamefont {V.~B.}\ \bibnamefont
  {Shenoy}},\ }\bibfield  {title} {\emph {\enquote {\bibinfo {title}
  {Topological Insulators in Amorphous Systems},}\ }}\href {\doibase
  10.1103/PhysRevLett.118.236402} {\bibfield  {journal} {\bibinfo  {journal}
  {Phys. Rev. Lett.}\ }\textbf {\bibinfo {volume} {118}},\ \bibinfo {pages}
  {236402} (\bibinfo {year} {2017})}\BibitemShut {NoStop}%
\bibitem [{\citenamefont {Mitchell}\ \emph {et~al.}(2018)\citenamefont
  {Mitchell}, \citenamefont {Nash}, \citenamefont {Hexner}, \citenamefont
  {Turner},\ and\ \citenamefont {Irvine}}]{MitchellNP2018}%
  \BibitemOpen
  \bibfield  {author} {\bibinfo {author} {\bibfnamefont {N.~P.}\ \bibnamefont
  {Mitchell}}, \bibinfo {author} {\bibfnamefont {L.~M.}\ \bibnamefont {Nash}},
  \bibinfo {author} {\bibfnamefont {D.}~\bibnamefont {Hexner}}, \bibinfo
  {author} {\bibfnamefont {A.~M.}\ \bibnamefont {Turner}}, \ and\ \bibinfo
  {author} {\bibfnamefont {W.~T.~M.}\ \bibnamefont {Irvine}},\ }\bibfield
  {title} {\emph {\enquote {\bibinfo {title} {Amorphous topological insulators
  constructed from random point sets},}\ }}\href {\doibase
  10.1038/s41567-017-0024-5} {\bibfield  {journal} {\bibinfo  {journal} {Nature
  Phys}\ }\textbf {\bibinfo {volume} {14}},\ \bibinfo {pages} {380} (\bibinfo
  {year} {2018})}\BibitemShut {NoStop}%
\bibitem [{\citenamefont {P{\"o}yh{\"o}nen}\ \emph {et~al.}(2018)\citenamefont
  {P{\"o}yh{\"o}nen}, \citenamefont {Sahlberg}, \citenamefont {Weststr{\"o}m},\
  and\ \citenamefont {Ojanen}}]{PoyhonenNC2018}%
  \BibitemOpen
  \bibfield  {author} {\bibinfo {author} {\bibfnamefont {K.}~\bibnamefont
  {P{\"o}yh{\"o}nen}}, \bibinfo {author} {\bibfnamefont {I.}~\bibnamefont
  {Sahlberg}}, \bibinfo {author} {\bibfnamefont {A.}~\bibnamefont
  {Weststr{\"o}m}}, \ and\ \bibinfo {author} {\bibfnamefont {T.}~\bibnamefont
  {Ojanen}},\ }\bibfield  {title} {\emph {\enquote {\bibinfo {title} {Amorphous
  topological superconductivity in a Shiba glass},}\ }}\href {\doibase
  10.1038/s41467-018-04532-x} {\bibfield  {journal} {\bibinfo  {journal} {Nat
  Commun}\ }\textbf {\bibinfo {volume} {9}},\ \bibinfo {pages} {2103} (\bibinfo
  {year} {2018})}\BibitemShut {NoStop}%
\bibitem [{\citenamefont {Sahlberg}\ \emph {et~al.}(2020)\citenamefont
  {Sahlberg}, \citenamefont {Weststr\"om}, \citenamefont {P\"oyh\"onen},\ and\
  \citenamefont {Ojanen}}]{SahlbergPRR2020}%
  \BibitemOpen
  \bibfield  {author} {\bibinfo {author} {\bibfnamefont {I.}~\bibnamefont
  {Sahlberg}}, \bibinfo {author} {\bibfnamefont {A.}~\bibnamefont
  {Weststr\"om}}, \bibinfo {author} {\bibfnamefont {K.}~\bibnamefont
  {P\"oyh\"onen}}, \ and\ \bibinfo {author} {\bibfnamefont {T.}~\bibnamefont
  {Ojanen}},\ }\bibfield  {title} {\emph {\enquote {\bibinfo {title}
  {Topological phase transitions in glassy quantum matter},}\ }}\href {\doibase
  10.1103/PhysRevResearch.2.013053} {\bibfield  {journal} {\bibinfo  {journal}
  {Phys. Rev. Res.}\ }\textbf {\bibinfo {volume} {2}},\ \bibinfo {pages}
  {013053} (\bibinfo {year} {2020})}\BibitemShut {NoStop}%
\bibitem [{\citenamefont {Yang}\ \emph {et~al.}(2019)\citenamefont {Yang},
  \citenamefont {Qin}, \citenamefont {Deng}, \citenamefont {Duan},\ and\
  \citenamefont {Xu}}]{YangAMPRL2019}%
  \BibitemOpen
  \bibfield  {author} {\bibinfo {author} {\bibfnamefont {Y.-B.}\ \bibnamefont
  {Yang}}, \bibinfo {author} {\bibfnamefont {T.}~\bibnamefont {Qin}}, \bibinfo
  {author} {\bibfnamefont {D.-L.}\ \bibnamefont {Deng}}, \bibinfo {author}
  {\bibfnamefont {L.-M.}\ \bibnamefont {Duan}}, \ and\ \bibinfo {author}
  {\bibfnamefont {Y.}~\bibnamefont {Xu}},\ }\bibfield  {title} {\emph {\enquote
  {\bibinfo {title} {Topological Amorphous Metals},}\ }}\href {\doibase
  10.1103/PhysRevLett.123.076401} {\bibfield  {journal} {\bibinfo  {journal}
  {Phys. Rev. Lett.}\ }\textbf {\bibinfo {volume} {123}},\ \bibinfo {pages}
  {076401} (\bibinfo {year} {2019})}\BibitemShut {NoStop}%
\bibitem [{\citenamefont {Costa}\ \emph {et~al.}(2019)\citenamefont {Costa},
  \citenamefont {Schleder}, \citenamefont {Buongiorno~Nardelli}, \citenamefont
  {Lewenkopf},\ and\ \citenamefont {Fazzio}}]{CostaNL2019}%
  \BibitemOpen
  \bibfield  {author} {\bibinfo {author} {\bibfnamefont {M.}~\bibnamefont
  {Costa}}, \bibinfo {author} {\bibfnamefont {G.~R.}\ \bibnamefont {Schleder}},
  \bibinfo {author} {\bibfnamefont {M.}~\bibnamefont {Buongiorno~Nardelli}},
  \bibinfo {author} {\bibfnamefont {C.}~\bibnamefont {Lewenkopf}}, \ and\
  \bibinfo {author} {\bibfnamefont {A.}~\bibnamefont {Fazzio}},\ }\bibfield
  {title} {\emph {\enquote {\bibinfo {title} {Toward Realistic Amorphous
  Topological Insulators},}\ }}\href {\doibase 10.1021/acs.nanolett.9b03881}
  {\bibfield  {journal} {\bibinfo  {journal} {Nano Lett.}\ }\textbf {\bibinfo
  {volume} {19}},\ \bibinfo {pages} {8941} (\bibinfo {year}
  {2019})}\BibitemShut {NoStop}%
\bibitem [{\citenamefont {Marsal}\ \emph {et~al.}(2020)\citenamefont {Marsal},
  \citenamefont {Varjas},\ and\ \citenamefont {Grushin}}]{MarsalPNAS2020}%
  \BibitemOpen
  \bibfield  {author} {\bibinfo {author} {\bibfnamefont {Q.}~\bibnamefont
  {Marsal}}, \bibinfo {author} {\bibfnamefont {D.}~\bibnamefont {Varjas}}, \
  and\ \bibinfo {author} {\bibfnamefont {A.~G.}\ \bibnamefont {Grushin}},\
  }\bibfield  {title} {\emph {\enquote {\bibinfo {title} {Topological
  Weaire–Thorpe models of amorphous matter},}\ }}\href {\doibase
  10.1073/pnas.2007384117} {\bibfield  {journal} {\bibinfo  {journal} {Proc.
  Natl. Acad. Sci.}\ }\textbf {\bibinfo {volume} {117}},\ \bibinfo {pages}
  {30260} (\bibinfo {year} {2020})}\BibitemShut {NoStop}%
\bibitem [{\citenamefont {Zhou}\ \emph {et~al.}(2020)\citenamefont {Zhou},
  \citenamefont {Liu}, \citenamefont {Ren}, \citenamefont {Yang}, \citenamefont
  {Xue}, \citenamefont {Bi}, \citenamefont {Deng}, \citenamefont {Chong},\ and\
  \citenamefont {Zhang}}]{ZhouLSA2020}%
  \BibitemOpen
  \bibfield  {author} {\bibinfo {author} {\bibfnamefont {P.}~\bibnamefont
  {Zhou}}, \bibinfo {author} {\bibfnamefont {G.-G.}\ \bibnamefont {Liu}},
  \bibinfo {author} {\bibfnamefont {X.}~\bibnamefont {Ren}}, \bibinfo {author}
  {\bibfnamefont {Y.}~\bibnamefont {Yang}}, \bibinfo {author} {\bibfnamefont
  {H.}~\bibnamefont {Xue}}, \bibinfo {author} {\bibfnamefont {L.}~\bibnamefont
  {Bi}}, \bibinfo {author} {\bibfnamefont {L.}~\bibnamefont {Deng}}, \bibinfo
  {author} {\bibfnamefont {Y.}~\bibnamefont {Chong}}, \ and\ \bibinfo {author}
  {\bibfnamefont {B.}~\bibnamefont {Zhang}},\ }\bibfield  {title} {\emph
  {\enquote {\bibinfo {title} {Photonic amorphous topological insulator},}\
  }}\href {\doibase 10.1038/s41377-020-00368-7} {\bibfield  {journal} {\bibinfo
   {journal} {Light Sci Appl}\ }\textbf {\bibinfo {volume} {9}},\ \bibinfo
  {pages} {133} (\bibinfo {year} {2020})}\BibitemShut {NoStop}%
\bibitem [{\citenamefont {Wang}\ \emph {et~al.}(2021)\citenamefont {Wang},
  \citenamefont {Yang}, \citenamefont {Dai},\ and\ \citenamefont
  {Xu}}]{WangPRLStructural2021}%
  \BibitemOpen
  \bibfield  {author} {\bibinfo {author} {\bibfnamefont {J.-H.}\ \bibnamefont
  {Wang}}, \bibinfo {author} {\bibfnamefont {Y.-B.}\ \bibnamefont {Yang}},
  \bibinfo {author} {\bibfnamefont {N.}~\bibnamefont {Dai}}, \ and\ \bibinfo
  {author} {\bibfnamefont {Y.}~\bibnamefont {Xu}},\ }\bibfield  {title} {\emph
  {\enquote {\bibinfo {title} {Structural-Disorder-Induced Second-Order
  Topological Insulators in Three Dimensions},}\ }}\href {\doibase
  10.1103/PhysRevLett.126.206404} {\bibfield  {journal} {\bibinfo  {journal}
  {Phys. Rev. Lett.}\ }\textbf {\bibinfo {volume} {126}},\ \bibinfo {pages}
  {206404} (\bibinfo {year} {2021})}\BibitemShut {NoStop}%
\bibitem [{\citenamefont {Corbae}\ \emph {et~al.}(2021)\citenamefont {Corbae},
  \citenamefont {Hellman},\ and\ \citenamefont {Griffin}}]{CorbaePRB2021}%
  \BibitemOpen
  \bibfield  {author} {\bibinfo {author} {\bibfnamefont {P.}~\bibnamefont
  {Corbae}}, \bibinfo {author} {\bibfnamefont {F.}~\bibnamefont {Hellman}}, \
  and\ \bibinfo {author} {\bibfnamefont {S.~M.}\ \bibnamefont {Griffin}},\
  }\bibfield  {title} {\emph {\enquote {\bibinfo {title} {Structural
  disorder-driven topological phase transition in noncentrosymmetric BiTeI},}\
  }}\href {\doibase 10.1103/PhysRevB.103.214203} {\bibfield  {journal}
  {\bibinfo  {journal} {Phys. Rev. B}\ }\textbf {\bibinfo {volume} {103}},\
  \bibinfo {pages} {214203} (\bibinfo {year} {2021})}\BibitemShut {NoStop}%
\bibitem [{\citenamefont {Li}\ \emph {et~al.}(2021)\citenamefont {Li},
  \citenamefont {Wang}, \citenamefont {Yang},\ and\ \citenamefont
  {Xu}}]{LiPRLTyd2021}%
  \BibitemOpen
  \bibfield  {author} {\bibinfo {author} {\bibfnamefont {K.}~\bibnamefont
  {Li}}, \bibinfo {author} {\bibfnamefont {J.-H.}\ \bibnamefont {Wang}},
  \bibinfo {author} {\bibfnamefont {Y.-B.}\ \bibnamefont {Yang}}, \ and\
  \bibinfo {author} {\bibfnamefont {Y.}~\bibnamefont {Xu}},\ }\bibfield
  {title} {\emph {\enquote {\bibinfo {title} {Symmetry-Protected Topological
  Phases in a Rydberg Glass},}\ }}\href {\doibase
  10.1103/PhysRevLett.127.263004} {\bibfield  {journal} {\bibinfo  {journal}
  {Phys. Rev. Lett.}\ }\textbf {\bibinfo {volume} {127}},\ \bibinfo {pages}
  {263004} (\bibinfo {year} {2021})}\BibitemShut {NoStop}%
\bibitem [{\citenamefont {Wang}\ \emph {et~al.}(2022)\citenamefont {Wang},
  \citenamefont {Cheng}, \citenamefont {Liu}, \citenamefont {Liu},\ and\
  \citenamefont {Huang}}]{WangAMIPRL2022}%
  \BibitemOpen
  \bibfield  {author} {\bibinfo {author} {\bibfnamefont {C.}~\bibnamefont
  {Wang}}, \bibinfo {author} {\bibfnamefont {T.}~\bibnamefont {Cheng}},
  \bibinfo {author} {\bibfnamefont {Z.}~\bibnamefont {Liu}}, \bibinfo {author}
  {\bibfnamefont {F.}~\bibnamefont {Liu}}, \ and\ \bibinfo {author}
  {\bibfnamefont {H.}~\bibnamefont {Huang}},\ }\bibfield  {title} {\emph
  {\enquote {\bibinfo {title} {Structural Amorphization-Induced Topological
  Order},}\ }}\href {\doibase 10.1103/PhysRevLett.128.056401} {\bibfield
  {journal} {\bibinfo  {journal} {Phys. Rev. Lett.}\ }\textbf {\bibinfo
  {volume} {128}},\ \bibinfo {pages} {056401} (\bibinfo {year}
  {2022})}\BibitemShut {NoStop}%
\bibitem [{\citenamefont {Le~Gallo}\ and\ \citenamefont
  {Sebastian}(2020)}]{legallo2020}%
  \BibitemOpen
  \bibfield  {author} {\bibinfo {author} {\bibfnamefont {M.}~\bibnamefont
  {Le~Gallo}}\ and\ \bibinfo {author} {\bibfnamefont {A.}~\bibnamefont
  {Sebastian}},\ }\bibfield  {title} {\emph {\enquote {\bibinfo {title} {An
  overview of phase-change memory device physics},}\ }}\href {\doibase
  10.1088/1361-6463/ab7794} {\bibfield  {journal} {\bibinfo  {journal} {Journal
  of Physics D: Applied Physics}\ }\textbf {\bibinfo {volume} {53}},\ \bibinfo
  {pages} {213002} (\bibinfo {year} {2020})}\BibitemShut {NoStop}%
\bibitem [{\citenamefont {Liu}\ \emph {et~al.}(2020)\citenamefont {Liu},
  \citenamefont {Yang}, \citenamefont {Ren}, \citenamefont {Xue}, \citenamefont
  {Lin}, \citenamefont {Hu}, \citenamefont {Sun}, \citenamefont {Peng},
  \citenamefont {Zhou}, \citenamefont {Chong},\ and\ \citenamefont
  {Zhang}}]{Liu2020}%
  \BibitemOpen
  \bibfield  {author} {\bibinfo {author} {\bibfnamefont {G.-G.}\ \bibnamefont
  {Liu}}, \bibinfo {author} {\bibfnamefont {Y.}~\bibnamefont {Yang}}, \bibinfo
  {author} {\bibfnamefont {X.}~\bibnamefont {Ren}}, \bibinfo {author}
  {\bibfnamefont {H.}~\bibnamefont {Xue}}, \bibinfo {author} {\bibfnamefont
  {X.}~\bibnamefont {Lin}}, \bibinfo {author} {\bibfnamefont {Y.-H.}\
  \bibnamefont {Hu}}, \bibinfo {author} {\bibfnamefont {H.-x.}\ \bibnamefont
  {Sun}}, \bibinfo {author} {\bibfnamefont {B.}~\bibnamefont {Peng}}, \bibinfo
  {author} {\bibfnamefont {P.}~\bibnamefont {Zhou}}, \bibinfo {author}
  {\bibfnamefont {Y.}~\bibnamefont {Chong}}, \ and\ \bibinfo {author}
  {\bibfnamefont {B.}~\bibnamefont {Zhang}},\ }\bibfield  {title} {\emph
  {\enquote {\bibinfo {title} {Topological Anderson Insulator in Disordered
  Photonic Crystals},}\ }}\href {\doibase 10.1103/PhysRevLett.125.133603}
  {\bibfield  {journal} {\bibinfo  {journal} {Phys. Rev. Lett.}\ }\textbf
  {\bibinfo {volume} {125}},\ \bibinfo {pages} {133603} (\bibinfo {year}
  {2020})}\BibitemShut {NoStop}%
\bibitem [{\citenamefont {Jia}\ \emph {et~al.}(2023)\citenamefont {Jia},
  \citenamefont {Seclì}, \citenamefont {Avdoshkin}, \citenamefont {Redjem},
  \citenamefont {Dresselhaus}, \citenamefont {Moore},\ and\ \citenamefont
  {Kanté}}]{Jia2022}%
  \BibitemOpen
  \bibfield  {author} {\bibinfo {author} {\bibfnamefont {Z.}~\bibnamefont
  {Jia}}, \bibinfo {author} {\bibfnamefont {M.}~\bibnamefont {Seclì}},
  \bibinfo {author} {\bibfnamefont {A.}~\bibnamefont {Avdoshkin}}, \bibinfo
  {author} {\bibfnamefont {W.}~\bibnamefont {Redjem}}, \bibinfo {author}
  {\bibfnamefont {E.}~\bibnamefont {Dresselhaus}}, \bibinfo {author}
  {\bibfnamefont {J.}~\bibnamefont {Moore}}, \ and\ \bibinfo {author}
  {\bibfnamefont {B.}~\bibnamefont {Kanté}},\ }\bibfield  {title} {\emph
  {\enquote {\bibinfo {title} {Disordered topological graphs enhancing
  nonlinear phenomena},}\ }}\href {\doibase 10.1126/sciadv.adf9330} {\bibfield
  {journal} {\bibinfo  {journal} {Sci. Adv.}\ }\textbf {\bibinfo {volume}
  {9}},\ \bibinfo {pages} {eadf9330} (\bibinfo {year} {2023})}\BibitemShut
  {NoStop}%
\bibitem [{\citenamefont {Corbae}\ \emph
  {et~al.}(2023{\natexlab{b}})\citenamefont {Corbae}, \citenamefont {Ciocys},
  \citenamefont {Varjas}, \citenamefont {Kennedy}, \citenamefont {Zeltmann},
  \citenamefont {Molina-Ruiz}, \citenamefont {Griffin}, \citenamefont
  {Jozwiak}, \citenamefont {Chen}, \citenamefont {Wang}, \citenamefont {Minor},
  \citenamefont {Scott}, \citenamefont {Grushin}, \citenamefont {Lanzara},\
  and\ \citenamefont {Hellman}}]{CorbaeNatMater2023}%
  \BibitemOpen
  \bibfield  {author} {\bibinfo {author} {\bibfnamefont {P.}~\bibnamefont
  {Corbae}}, \bibinfo {author} {\bibfnamefont {S.}~\bibnamefont {Ciocys}},
  \bibinfo {author} {\bibfnamefont {D.}~\bibnamefont {Varjas}}, \bibinfo
  {author} {\bibfnamefont {E.}~\bibnamefont {Kennedy}}, \bibinfo {author}
  {\bibfnamefont {S.}~\bibnamefont {Zeltmann}}, \bibinfo {author}
  {\bibfnamefont {M.}~\bibnamefont {Molina-Ruiz}}, \bibinfo {author}
  {\bibfnamefont {S.~M.}\ \bibnamefont {Griffin}}, \bibinfo {author}
  {\bibfnamefont {C.}~\bibnamefont {Jozwiak}}, \bibinfo {author} {\bibfnamefont
  {Z.}~\bibnamefont {Chen}}, \bibinfo {author} {\bibfnamefont {L.-W.}\
  \bibnamefont {Wang}}, \bibinfo {author} {\bibfnamefont {A.~M.}\ \bibnamefont
  {Minor}}, \bibinfo {author} {\bibfnamefont {M.}~\bibnamefont {Scott}},
  \bibinfo {author} {\bibfnamefont {A.~G.}\ \bibnamefont {Grushin}}, \bibinfo
  {author} {\bibfnamefont {A.}~\bibnamefont {Lanzara}}, \ and\ \bibinfo
  {author} {\bibfnamefont {F.}~\bibnamefont {Hellman}},\ }\bibfield  {title}
  {\emph {\enquote {\bibinfo {title} {Observation of spin-momentum locked
  surface states in amorphous Bi$_2$Se$_3$},}\ }}\href {\doibase
  10.1038/s41563-022-01458-0} {\bibfield  {journal} {\bibinfo  {journal} {Nat.
  Mater.}\ }\textbf {\bibinfo {volume} {22}},\ \bibinfo {pages} {200} (\bibinfo
  {year} {2023}{\natexlab{b}})}\BibitemShut {NoStop}%
\bibitem [{\citenamefont {Ciocys}\ \emph {et~al.}(2024)\citenamefont {Ciocys},
  \citenamefont {Marsal}, \citenamefont {Corbae}, \citenamefont {Varjas},
  \citenamefont {Kennedy}, \citenamefont {Scott}, \citenamefont {Hellman},
  \citenamefont {Grushin},\ and\ \citenamefont {Lanzara}}]{CiocysNatComm2024}%
  \BibitemOpen
  \bibfield  {author} {\bibinfo {author} {\bibfnamefont {S.~T.}\ \bibnamefont
  {Ciocys}}, \bibinfo {author} {\bibfnamefont {Q.}~\bibnamefont {Marsal}},
  \bibinfo {author} {\bibfnamefont {P.}~\bibnamefont {Corbae}}, \bibinfo
  {author} {\bibfnamefont {D.}~\bibnamefont {Varjas}}, \bibinfo {author}
  {\bibfnamefont {E.}~\bibnamefont {Kennedy}}, \bibinfo {author} {\bibfnamefont
  {M.}~\bibnamefont {Scott}}, \bibinfo {author} {\bibfnamefont
  {F.}~\bibnamefont {Hellman}}, \bibinfo {author} {\bibfnamefont {A.~G.}\
  \bibnamefont {Grushin}}, \ and\ \bibinfo {author} {\bibfnamefont
  {A.}~\bibnamefont {Lanzara}},\ }\bibfield  {title} {\emph {\enquote {\bibinfo
  {title} {Establishing coherent momentum-space electronic states in locally
  ordered materials},}\ }}\href {\doibase 10.1038/s41467-024-51953-y}
  {\bibfield  {journal} {\bibinfo  {journal} {Nat. Commun.}\ }\textbf {\bibinfo
  {volume} {15}},\ \bibinfo {pages} {8141} (\bibinfo {year}
  {2024})}\BibitemShut {NoStop}%
\bibitem [{\citenamefont {Toh}\ \emph {et~al.}(2020)\citenamefont {Toh},
  \citenamefont {Zhang}, \citenamefont {Lin}, \citenamefont {Mayorov},
  \citenamefont {Wang}, \citenamefont {Orofeo}, \citenamefont {Ferry},
  \citenamefont {Andersen}, \citenamefont {Kakenov}, \citenamefont {Guo},
  \citenamefont {Abidi}, \citenamefont {Sims}, \citenamefont {Suenaga},
  \citenamefont {Pantelides},\ and\ \citenamefont
  {{\"O}zyilmaz}}]{TohNature2020}%
  \BibitemOpen
  \bibfield  {author} {\bibinfo {author} {\bibfnamefont {C.-T.}\ \bibnamefont
  {Toh}}, \bibinfo {author} {\bibfnamefont {H.}~\bibnamefont {Zhang}}, \bibinfo
  {author} {\bibfnamefont {J.}~\bibnamefont {Lin}}, \bibinfo {author}
  {\bibfnamefont {A.~S.}\ \bibnamefont {Mayorov}}, \bibinfo {author}
  {\bibfnamefont {Y.-P.}\ \bibnamefont {Wang}}, \bibinfo {author}
  {\bibfnamefont {C.~M.}\ \bibnamefont {Orofeo}}, \bibinfo {author}
  {\bibfnamefont {D.~B.}\ \bibnamefont {Ferry}}, \bibinfo {author}
  {\bibfnamefont {H.}~\bibnamefont {Andersen}}, \bibinfo {author}
  {\bibfnamefont {N.}~\bibnamefont {Kakenov}}, \bibinfo {author} {\bibfnamefont
  {Z.}~\bibnamefont {Guo}}, \bibinfo {author} {\bibfnamefont {I.~H.}\
  \bibnamefont {Abidi}}, \bibinfo {author} {\bibfnamefont {H.}~\bibnamefont
  {Sims}}, \bibinfo {author} {\bibfnamefont {K.}~\bibnamefont {Suenaga}},
  \bibinfo {author} {\bibfnamefont {S.~T.}\ \bibnamefont {Pantelides}}, \ and\
  \bibinfo {author} {\bibfnamefont {B.}~\bibnamefont {{\"O}zyilmaz}},\
  }\bibfield  {title} {\emph {\enquote {\bibinfo {title} {Synthesis and
  properties of free-standing monolayer amorphous carbon},}\ }}\href {\doibase
  10.1038/s41586-019-1871-2} {\bibfield  {journal} {\bibinfo  {journal}
  {Nature}\ }\textbf {\bibinfo {volume} {577}},\ \bibinfo {pages} {199}
  (\bibinfo {year} {2020})}\BibitemShut {NoStop}%
\bibitem [{\citenamefont {Ghosh}\ \emph {et~al.}(2024)\citenamefont {Ghosh},
  \citenamefont {Arouca},\ and\ \citenamefont {Black-Schaffer}}]{GhoshSL2024}%
  \BibitemOpen
  \bibfield  {author} {\bibinfo {author} {\bibfnamefont {A.~K.}\ \bibnamefont
  {Ghosh}}, \bibinfo {author} {\bibfnamefont {R.}~\bibnamefont {Arouca}}, \
  and\ \bibinfo {author} {\bibfnamefont {A.~M.}\ \bibnamefont
  {Black-Schaffer}},\ }\bibfield  {title} {\emph {\enquote {\bibinfo {title}
  {Local and energy-resolved topological invariants for Floquet systems},}\
  }}\href {\doibase 10.1103/PhysRevB.110.245306} {\bibfield  {journal}
  {\bibinfo  {journal} {Phys. Rev. B}\ }\textbf {\bibinfo {volume} {110}},\
  \bibinfo {pages} {245306} (\bibinfo {year} {2024})}\BibitemShut {NoStop}%
\bibitem [{\citenamefont {Krasheninnikov}\ \emph {et~al.}(2009)\citenamefont
  {Krasheninnikov}, \citenamefont {Lehtinen}, \citenamefont {Foster},
  \citenamefont {Pyykk\"o},\ and\ \citenamefont
  {Nieminen}}]{KrasheninnikovPRL2009}%
  \BibitemOpen
  \bibfield  {author} {\bibinfo {author} {\bibfnamefont {A.~V.}\ \bibnamefont
  {Krasheninnikov}}, \bibinfo {author} {\bibfnamefont {P.~O.}\ \bibnamefont
  {Lehtinen}}, \bibinfo {author} {\bibfnamefont {A.~S.}\ \bibnamefont
  {Foster}}, \bibinfo {author} {\bibfnamefont {P.}~\bibnamefont {Pyykk\"o}}, \
  and\ \bibinfo {author} {\bibfnamefont {R.~M.}\ \bibnamefont {Nieminen}},\
  }\bibfield  {title} {\emph {\enquote {\bibinfo {title} {Embedding
  Transition-Metal Atoms in Graphene: Structure, Bonding, and Magnetism},}\
  }}\href {\doibase 10.1103/PhysRevLett.102.126807} {\bibfield  {journal}
  {\bibinfo  {journal} {Phys. Rev. Lett.}\ }\textbf {\bibinfo {volume} {102}},\
  \bibinfo {pages} {126807} (\bibinfo {year} {2009})}\BibitemShut {NoStop}%
\bibitem [{\citenamefont {Zhou}\ \emph {et~al.}(2012)\citenamefont {Zhou},
  \citenamefont {Kapetanakis}, \citenamefont {Prange}, \citenamefont
  {Pantelides}, \citenamefont {Pennycook},\ and\ \citenamefont
  {Idrobo}}]{ZhouPRL2012}%
  \BibitemOpen
  \bibfield  {author} {\bibinfo {author} {\bibfnamefont {W.}~\bibnamefont
  {Zhou}}, \bibinfo {author} {\bibfnamefont {M.~D.}\ \bibnamefont
  {Kapetanakis}}, \bibinfo {author} {\bibfnamefont {M.~P.}\ \bibnamefont
  {Prange}}, \bibinfo {author} {\bibfnamefont {S.~T.}\ \bibnamefont
  {Pantelides}}, \bibinfo {author} {\bibfnamefont {S.~J.}\ \bibnamefont
  {Pennycook}}, \ and\ \bibinfo {author} {\bibfnamefont {J.-C.}\ \bibnamefont
  {Idrobo}},\ }\bibfield  {title} {\emph {\enquote {\bibinfo {title} {Direct
  Determination of the Chemical Bonding of Individual Impurities in
  Graphene},}\ }}\href {\doibase 10.1103/PhysRevLett.109.206803} {\bibfield
  {journal} {\bibinfo  {journal} {Phys. Rev. Lett.}\ }\textbf {\bibinfo
  {volume} {109}},\ \bibinfo {pages} {206803} (\bibinfo {year}
  {2012})}\BibitemShut {NoStop}%
\bibitem [{\citenamefont {Wang}\ \emph {et~al.}(2012)\citenamefont {Wang},
  \citenamefont {Wang}, \citenamefont {Cheng}, \citenamefont {Li},
  \citenamefont {Yao}, \citenamefont {Zhang}, \citenamefont {Dong},
  \citenamefont {Wang}, \citenamefont {Schwingenschl{\"o}gl}, \citenamefont
  {Yang},\ and\ \citenamefont {Zhang}}]{WangNL2012}%
  \BibitemOpen
  \bibfield  {author} {\bibinfo {author} {\bibfnamefont {H.}~\bibnamefont
  {Wang}}, \bibinfo {author} {\bibfnamefont {Q.}~\bibnamefont {Wang}}, \bibinfo
  {author} {\bibfnamefont {Y.}~\bibnamefont {Cheng}}, \bibinfo {author}
  {\bibfnamefont {K.}~\bibnamefont {Li}}, \bibinfo {author} {\bibfnamefont
  {Y.}~\bibnamefont {Yao}}, \bibinfo {author} {\bibfnamefont {Q.}~\bibnamefont
  {Zhang}}, \bibinfo {author} {\bibfnamefont {C.}~\bibnamefont {Dong}},
  \bibinfo {author} {\bibfnamefont {P.}~\bibnamefont {Wang}}, \bibinfo {author}
  {\bibfnamefont {U.}~\bibnamefont {Schwingenschl{\"o}gl}}, \bibinfo {author}
  {\bibfnamefont {W.}~\bibnamefont {Yang}}, \ and\ \bibinfo {author}
  {\bibfnamefont {X.~X.}\ \bibnamefont {Zhang}},\ }\bibfield  {title} {\emph
  {\enquote {\bibinfo {title} {Doping Monolayer Graphene with Single Atom
  Substitutions},}\ }}\href {\doibase 10.1021/nl2031629} {\bibfield  {journal}
  {\bibinfo  {journal} {Nano Lett.}\ }\textbf {\bibinfo {volume} {12}},\
  \bibinfo {pages} {141} (\bibinfo {year} {2012})}\BibitemShut {NoStop}%
\bibitem [{\citenamefont {Ramasse}\ \emph {et~al.}(2013)\citenamefont
  {Ramasse}, \citenamefont {Seabourne}, \citenamefont {Kepaptsoglou},
  \citenamefont {Zan}, \citenamefont {Bangert},\ and\ \citenamefont
  {Scott}}]{RamasseNL2013}%
  \BibitemOpen
  \bibfield  {author} {\bibinfo {author} {\bibfnamefont {Q.~M.}\ \bibnamefont
  {Ramasse}}, \bibinfo {author} {\bibfnamefont {C.~R.}\ \bibnamefont
  {Seabourne}}, \bibinfo {author} {\bibfnamefont {D.-M.}\ \bibnamefont
  {Kepaptsoglou}}, \bibinfo {author} {\bibfnamefont {R.}~\bibnamefont {Zan}},
  \bibinfo {author} {\bibfnamefont {U.}~\bibnamefont {Bangert}}, \ and\
  \bibinfo {author} {\bibfnamefont {A.~J.}\ \bibnamefont {Scott}},\ }\bibfield
  {title} {\emph {\enquote {\bibinfo {title} {Probing the Bonding and
  Electronic Structure of Single Atom Dopants in Graphene with Electron Energy
  Loss Spectroscopy},}\ }}\href {\doibase 10.1021/nl304187e} {\bibfield
  {journal} {\bibinfo  {journal} {Nano Lett.}\ }\textbf {\bibinfo {volume}
  {13}},\ \bibinfo {pages} {4989} (\bibinfo {year} {2013})}\BibitemShut
  {NoStop}%
\bibitem [{\citenamefont {Yan}\ \emph {et~al.}(2018)\citenamefont {Yan},
  \citenamefont {Jia},\ and\ \citenamefont {Yao}}]{YanCSR2018}%
  \BibitemOpen
  \bibfield  {author} {\bibinfo {author} {\bibfnamefont {X.}~\bibnamefont
  {Yan}}, \bibinfo {author} {\bibfnamefont {Y.}~\bibnamefont {Jia}}, \ and\
  \bibinfo {author} {\bibfnamefont {X.}~\bibnamefont {Yao}},\ }\bibfield
  {title} {\emph {\enquote {\bibinfo {title} {Defects on carbons for
  electrocatalytic oxygen reduction},}\ }}\href {\doibase 10.1039/C7CS00690J}
  {\bibfield  {journal} {\bibinfo  {journal} {Chem. Soc. Rev.}\ }\textbf
  {\bibinfo {volume} {47}},\ \bibinfo {pages} {7628} (\bibinfo {year}
  {2018})}\BibitemShut {NoStop}%
\bibitem [{\citenamefont {Kotakoski}\ \emph {et~al.}(2011)\citenamefont
  {Kotakoski}, \citenamefont {Krasheninnikov}, \citenamefont {Kaiser},\ and\
  \citenamefont {Meyer}}]{KotakoskiPRL2011}%
  \BibitemOpen
  \bibfield  {author} {\bibinfo {author} {\bibfnamefont {J.}~\bibnamefont
  {Kotakoski}}, \bibinfo {author} {\bibfnamefont {A.~V.}\ \bibnamefont
  {Krasheninnikov}}, \bibinfo {author} {\bibfnamefont {U.}~\bibnamefont
  {Kaiser}}, \ and\ \bibinfo {author} {\bibfnamefont {J.~C.}\ \bibnamefont
  {Meyer}},\ }\bibfield  {title} {\emph {\enquote {\bibinfo {title} {From Point
  Defects in Graphene to Two-Dimensional Amorphous Carbon},}\ }}\href {\doibase
  10.1103/PhysRevLett.106.105505} {\bibfield  {journal} {\bibinfo  {journal}
  {Phys. Rev. Lett.}\ }\textbf {\bibinfo {volume} {106}},\ \bibinfo {pages}
  {105505} (\bibinfo {year} {2011})}\BibitemShut {NoStop}%
\bibitem [{\citenamefont {Huang}\ \emph {et~al.}(2011)\citenamefont {Huang},
  \citenamefont {Ruiz-Vargas}, \citenamefont {van~der Zande}, \citenamefont
  {Whitney}, \citenamefont {Levendorf}, \citenamefont {Kevek}, \citenamefont
  {Garg}, \citenamefont {Alden}, \citenamefont {Hustedt}, \citenamefont {Zhu},
  \citenamefont {Park}, \citenamefont {McEuen},\ and\ \citenamefont
  {Muller}}]{HuangNP2011}%
  \BibitemOpen
  \bibfield  {author} {\bibinfo {author} {\bibfnamefont {P.~Y.}\ \bibnamefont
  {Huang}}, \bibinfo {author} {\bibfnamefont {C.~S.}\ \bibnamefont
  {Ruiz-Vargas}}, \bibinfo {author} {\bibfnamefont {A.~M.}\ \bibnamefont
  {van~der Zande}}, \bibinfo {author} {\bibfnamefont {W.~S.}\ \bibnamefont
  {Whitney}}, \bibinfo {author} {\bibfnamefont {M.~P.}\ \bibnamefont
  {Levendorf}}, \bibinfo {author} {\bibfnamefont {J.~W.}\ \bibnamefont
  {Kevek}}, \bibinfo {author} {\bibfnamefont {S.}~\bibnamefont {Garg}},
  \bibinfo {author} {\bibfnamefont {J.~S.}\ \bibnamefont {Alden}}, \bibinfo
  {author} {\bibfnamefont {C.~J.}\ \bibnamefont {Hustedt}}, \bibinfo {author}
  {\bibfnamefont {Y.}~\bibnamefont {Zhu}}, \bibinfo {author} {\bibfnamefont
  {J.}~\bibnamefont {Park}}, \bibinfo {author} {\bibfnamefont {P.~L.}\
  \bibnamefont {McEuen}}, \ and\ \bibinfo {author} {\bibfnamefont {D.~A.}\
  \bibnamefont {Muller}},\ }\bibfield  {title} {\emph {\enquote {\bibinfo
  {title} {Grains and grain boundaries in single-layer graphene atomic
  patchwork quilts},}\ }}\href {\doibase 10.1038/nature09718} {\bibfield
  {journal} {\bibinfo  {journal} {Nature}\ }\textbf {\bibinfo {volume} {469}},\
  \bibinfo {pages} {389} (\bibinfo {year} {2011})}\BibitemShut {NoStop}%
\bibitem [{\citenamefont {Kim}\ \emph {et~al.}(2011)\citenamefont {Kim},
  \citenamefont {Lee}, \citenamefont {Regan}, \citenamefont {Kisielowski},
  \citenamefont {Crommie},\ and\ \citenamefont {Zettl}}]{KwanpyoACSNano2011}%
  \BibitemOpen
  \bibfield  {author} {\bibinfo {author} {\bibfnamefont {K.}~\bibnamefont
  {Kim}}, \bibinfo {author} {\bibfnamefont {Z.}~\bibnamefont {Lee}}, \bibinfo
  {author} {\bibfnamefont {W.}~\bibnamefont {Regan}}, \bibinfo {author}
  {\bibfnamefont {C.}~\bibnamefont {Kisielowski}}, \bibinfo {author}
  {\bibfnamefont {M.~F.}\ \bibnamefont {Crommie}}, \ and\ \bibinfo {author}
  {\bibfnamefont {A.}~\bibnamefont {Zettl}},\ }\bibfield  {title} {\emph
  {\enquote {\bibinfo {title} {Grain Boundary Mapping in Polycrystalline
  Graphene},}\ }}\href {\doibase 10.1021/nn1033423} {\bibfield  {journal}
  {\bibinfo  {journal} {ACS Nano}\ }\textbf {\bibinfo {volume} {5}},\ \bibinfo
  {pages} {2142} (\bibinfo {year} {2011})}\BibitemShut {NoStop}%
\bibitem [{\citenamefont {Yazyev}\ and\ \citenamefont
  {Chen}(2014)}]{YazyevNN2014}%
  \BibitemOpen
  \bibfield  {author} {\bibinfo {author} {\bibfnamefont {O.~V.}\ \bibnamefont
  {Yazyev}}\ and\ \bibinfo {author} {\bibfnamefont {Y.~P.}\ \bibnamefont
  {Chen}},\ }\bibfield  {title} {\emph {\enquote {\bibinfo {title}
  {Polycrystalline graphene and other two-dimensional materials},}\ }}\href
  {\doibase 10.1038/nnano.2014.166} {\bibfield  {journal} {\bibinfo  {journal}
  {Nature Nanotech}\ }\textbf {\bibinfo {volume} {9}},\ \bibinfo {pages} {755}
  (\bibinfo {year} {2014})}\BibitemShut {NoStop}%
\bibitem [{\citenamefont {Floquet}(1883)}]{FLoquetpaper1883}%
  \BibitemOpen
  \bibfield  {author} {\bibinfo {author} {\bibfnamefont {G.}~\bibnamefont
  {Floquet}},\ }\bibfield  {title} {\emph {\enquote {\bibinfo {title} {Sur les
  {\'e}quations diff{\'e}rentielles lin{\'e}aires {\`a} coefficients
  p{\'e}riodiques},}\ }}\href {\doibase 10.24033/asens.220} {\bibfield
  {journal} {\bibinfo  {journal} {Ann. Ecole Norm. Superieure}\ }\textbf
  {\bibinfo {volume} {12}},\ \bibinfo {pages} {47} (\bibinfo {year}
  {1883})}\BibitemShut {NoStop}%
\bibitem [{\citenamefont {Mikami}\ \emph {et~al.}(2016)\citenamefont {Mikami},
  \citenamefont {Kitamura}, \citenamefont {Yasuda}, \citenamefont {Tsuji},
  \citenamefont {Oka},\ and\ \citenamefont {Aoki}}]{MikamiBW2016}%
  \BibitemOpen
  \bibfield  {author} {\bibinfo {author} {\bibfnamefont {T.}~\bibnamefont
  {Mikami}}, \bibinfo {author} {\bibfnamefont {S.}~\bibnamefont {Kitamura}},
  \bibinfo {author} {\bibfnamefont {K.}~\bibnamefont {Yasuda}}, \bibinfo
  {author} {\bibfnamefont {N.}~\bibnamefont {Tsuji}}, \bibinfo {author}
  {\bibfnamefont {T.}~\bibnamefont {Oka}}, \ and\ \bibinfo {author}
  {\bibfnamefont {H.}~\bibnamefont {Aoki}},\ }\bibfield  {title} {\emph
  {\enquote {\bibinfo {title} {{Brillouin-Wigner theory for high-frequency
  expansion in periodically driven systems: Application to Floquet topological
  insulators}},}\ }}\href {\doibase 10.1103/PhysRevB.93.144307} {\bibfield
  {journal} {\bibinfo  {journal} {Phys. Rev. B}\ }\textbf {\bibinfo {volume}
  {93}},\ \bibinfo {pages} {144307} (\bibinfo {year} {2016})}\BibitemShut
  {NoStop}%
\bibitem [{\citenamefont {Merboldt}\ \emph {et~al.}(2025)\citenamefont
  {Merboldt}, \citenamefont {Sch{\"u}ler}, \citenamefont {Schmitt},
  \citenamefont {Bange}, \citenamefont {Bennecke}, \citenamefont {Gadge},
  \citenamefont {Pierz}, \citenamefont {Schumacher}, \citenamefont {Momeni},
  \citenamefont {Steil}, \citenamefont {Manmana}, \citenamefont {Sentef},
  \citenamefont {Reutzel},\ and\ \citenamefont {Mathias}}]{MerboldtNP2025}%
  \BibitemOpen
  \bibfield  {author} {\bibinfo {author} {\bibfnamefont {M.}~\bibnamefont
  {Merboldt}}, \bibinfo {author} {\bibfnamefont {M.}~\bibnamefont
  {Sch{\"u}ler}}, \bibinfo {author} {\bibfnamefont {D.}~\bibnamefont
  {Schmitt}}, \bibinfo {author} {\bibfnamefont {J.~P.}\ \bibnamefont {Bange}},
  \bibinfo {author} {\bibfnamefont {W.}~\bibnamefont {Bennecke}}, \bibinfo
  {author} {\bibfnamefont {K.}~\bibnamefont {Gadge}}, \bibinfo {author}
  {\bibfnamefont {K.}~\bibnamefont {Pierz}}, \bibinfo {author} {\bibfnamefont
  {H.~W.}\ \bibnamefont {Schumacher}}, \bibinfo {author} {\bibfnamefont
  {D.}~\bibnamefont {Momeni}}, \bibinfo {author} {\bibfnamefont
  {D.}~\bibnamefont {Steil}}, \bibinfo {author} {\bibfnamefont {S.~R.}\
  \bibnamefont {Manmana}}, \bibinfo {author} {\bibfnamefont {M.~A.}\
  \bibnamefont {Sentef}}, \bibinfo {author} {\bibfnamefont {M.}~\bibnamefont
  {Reutzel}}, \ and\ \bibinfo {author} {\bibfnamefont {S.}~\bibnamefont
  {Mathias}},\ }\bibfield  {title} {\emph {\enquote {\bibinfo {title}
  {Observation of Floquet states in graphene},}\ }}\href {\doibase
  10.1038/s41567-025-02889-7} {\bibfield  {journal} {\bibinfo  {journal} {Nat.
  Phys.}\ } (\bibinfo {year} {2025}),\ 10.1038/s41567-025-02889-7}\BibitemShut
  {NoStop}%
\bibitem [{sup()}]{supp}%
  \BibitemOpen
  \href@noop {} {}\bibinfo {note} {Supplemental Material at XXXXXXXXXXX for
  details of the topological invariant: LDLT decomposition, analysis of
  localizer gap, spatially resolved localizer gap and topological invariant,
  stable $\pi$-mode, and variance of the topological invariant, which includes
  Refs.~\cite{ChadhaPRB2024,BuenoLAA2007,CerjanJMP2023}.}\BibitemShut {Stop}%
\bibitem [{\citenamefont {Titum}\ \emph {et~al.}(2016)\citenamefont {Titum},
  \citenamefont {Berg}, \citenamefont {Rudner}, \citenamefont {Refael},\ and\
  \citenamefont {Lindner}}]{TitumPRX2016}%
  \BibitemOpen
  \bibfield  {author} {\bibinfo {author} {\bibfnamefont {P.}~\bibnamefont
  {Titum}}, \bibinfo {author} {\bibfnamefont {E.}~\bibnamefont {Berg}},
  \bibinfo {author} {\bibfnamefont {M.~S.}\ \bibnamefont {Rudner}}, \bibinfo
  {author} {\bibfnamefont {G.}~\bibnamefont {Refael}}, \ and\ \bibinfo {author}
  {\bibfnamefont {N.~H.}\ \bibnamefont {Lindner}},\ }\bibfield  {title} {\emph
  {\enquote {\bibinfo {title} {Anomalous Floquet-Anderson Insulator as a
  Nonadiabatic Quantized Charge Pump},}\ }}\href {\doibase
  10.1103/PhysRevX.6.021013} {\bibfield  {journal} {\bibinfo  {journal} {Phys.
  Rev. X}\ }\textbf {\bibinfo {volume} {6}},\ \bibinfo {pages} {021013}
  (\bibinfo {year} {2016})}\BibitemShut {NoStop}%
\bibitem [{\citenamefont {Peralta~Gavensky}\ \emph {et~al.}(2025)\citenamefont
  {Peralta~Gavensky}, \citenamefont {Usaj},\ and\ \citenamefont
  {Goldman}}]{GavenskyPRX2025}%
  \BibitemOpen
  \bibfield  {author} {\bibinfo {author} {\bibfnamefont {L.}~\bibnamefont
  {Peralta~Gavensky}}, \bibinfo {author} {\bibfnamefont {G.}~\bibnamefont
  {Usaj}}, \ and\ \bibinfo {author} {\bibfnamefont {N.}~\bibnamefont
  {Goldman}},\ }\bibfield  {title} {\emph {\enquote {\bibinfo {title}
  {St\ifmmode \check{r}\else \v{r}\fi{}eda Formula for Floquet Systems:
  Topological Invariants and Quantized Anomalies from Ces\`aro Summation},}\
  }}\href {\doibase 10.1103/b3pw-my97} {\bibfield  {journal} {\bibinfo
  {journal} {Phys. Rev. X}\ }\textbf {\bibinfo {volume} {15}},\ \bibinfo
  {pages} {031067} (\bibinfo {year} {2025})}\BibitemShut {NoStop}%
\bibitem [{\citenamefont {Loring}(2015)}]{LoringAnnPhys2015}%
  \BibitemOpen
  \bibfield  {author} {\bibinfo {author} {\bibfnamefont {T.~A.}\ \bibnamefont
  {Loring}},\ }\bibfield  {title} {\emph {\enquote {\bibinfo {title} {K-theory
  and pseudospectra for topological insulators},}\ }}\href {\doibase
  https://doi.org/10.1016/j.aop.2015.02.031} {\bibfield  {journal} {\bibinfo
  {journal} {Ann. Phys.}\ }\textbf {\bibinfo {volume} {356}},\ \bibinfo {pages}
  {383} (\bibinfo {year} {2015})}\BibitemShut {NoStop}%
\bibitem [{\citenamefont {Loring}\ and\ \citenamefont
  {Schulz‑Baldes}(2017)}]{loring2017finitevolume}%
  \BibitemOpen
  \bibfield  {author} {\bibinfo {author} {\bibfnamefont {T.~A.}\ \bibnamefont
  {Loring}}\ and\ \bibinfo {author} {\bibfnamefont {H.}~\bibnamefont
  {Schulz‑Baldes}},\ }\bibfield  {title} {\emph {\enquote {\bibinfo {title}
  {Finite volume calculation of $K$-theory invariants},}\ }}\href
  {http://nyjm.albany.edu/j/2017/23-48.html} {\bibfield  {journal} {\bibinfo
  {journal} {New York J. Math.}\ }\textbf {\bibinfo {volume} {23}},\ \bibinfo
  {pages} {1111} (\bibinfo {year} {2017})}\BibitemShut {NoStop}%
\bibitem [{\citenamefont {Loring}()}]{loring2019guide}%
  \BibitemOpen
  \bibfield  {author} {\bibinfo {author} {\bibfnamefont {T.~A.}\ \bibnamefont
  {Loring}},\ }\href@noop {} {\enquote {\bibinfo {title} {A Guide to the Bott
  Index and Localizer Index},}\ }\Eprint
  {http://arxiv.org/abs/1907.11791}{arXiv:1907.11791}\BibitemShut {NoStop}%
\bibitem [{\citenamefont {Nathan}\ \emph {et~al.}(2017)\citenamefont {Nathan},
  \citenamefont {Rudner}, \citenamefont {Lindner}, \citenamefont {Berg},\ and\
  \citenamefont {Refael}}]{NathanPRL2017}%
  \BibitemOpen
  \bibfield  {author} {\bibinfo {author} {\bibfnamefont {F.}~\bibnamefont
  {Nathan}}, \bibinfo {author} {\bibfnamefont {M.~S.}\ \bibnamefont {Rudner}},
  \bibinfo {author} {\bibfnamefont {N.~H.}\ \bibnamefont {Lindner}}, \bibinfo
  {author} {\bibfnamefont {E.}~\bibnamefont {Berg}}, \ and\ \bibinfo {author}
  {\bibfnamefont {G.}~\bibnamefont {Refael}},\ }\bibfield  {title} {\emph
  {\enquote {\bibinfo {title} {Quantized Magnetization Density in Periodically
  Driven Systems},}\ }}\href {\doibase 10.1103/PhysRevLett.119.186801}
  {\bibfield  {journal} {\bibinfo  {journal} {Phys. Rev. Lett.}\ }\textbf
  {\bibinfo {volume} {119}},\ \bibinfo {pages} {186801} (\bibinfo {year}
  {2017})}\BibitemShut {NoStop}%
\bibitem [{\citenamefont {Wang}\ \emph {et~al.}(2013)\citenamefont {Wang},
  \citenamefont {Steinberg}, \citenamefont {Jarillo-Herrero},\ and\
  \citenamefont {Gedik}}]{WangScience2013}%
  \BibitemOpen
  \bibfield  {author} {\bibinfo {author} {\bibfnamefont {Y.~H.}\ \bibnamefont
  {Wang}}, \bibinfo {author} {\bibfnamefont {H.}~\bibnamefont {Steinberg}},
  \bibinfo {author} {\bibfnamefont {P.}~\bibnamefont {Jarillo-Herrero}}, \ and\
  \bibinfo {author} {\bibfnamefont {N.}~\bibnamefont {Gedik}},\ }\bibfield
  {title} {\emph {\enquote {\bibinfo {title} {Observation of Floquet-Bloch
  States on the Surface of a Topological Insulator},}\ }}\href {\doibase
  10.1126/science.1239834} {\bibfield  {journal} {\bibinfo  {journal}
  {Science}\ }\textbf {\bibinfo {volume} {342}},\ \bibinfo {pages} {453}
  (\bibinfo {year} {2013})}\BibitemShut {NoStop}%
\bibitem [{\citenamefont {Zhou}\ \emph {et~al.}(2023)\citenamefont {Zhou},
  \citenamefont {Bao}, \citenamefont {Fan}, \citenamefont {Zhou}, \citenamefont
  {Gao}, \citenamefont {Zhong}, \citenamefont {Lin}, \citenamefont {Liu},
  \citenamefont {Yu}, \citenamefont {Tang}, \citenamefont {Meng}, \citenamefont
  {Duan},\ and\ \citenamefont {Zhou}}]{ZhouNature2023}%
  \BibitemOpen
  \bibfield  {author} {\bibinfo {author} {\bibfnamefont {S.}~\bibnamefont
  {Zhou}}, \bibinfo {author} {\bibfnamefont {C.}~\bibnamefont {Bao}}, \bibinfo
  {author} {\bibfnamefont {B.}~\bibnamefont {Fan}}, \bibinfo {author}
  {\bibfnamefont {H.}~\bibnamefont {Zhou}}, \bibinfo {author} {\bibfnamefont
  {Q.}~\bibnamefont {Gao}}, \bibinfo {author} {\bibfnamefont {H.}~\bibnamefont
  {Zhong}}, \bibinfo {author} {\bibfnamefont {T.}~\bibnamefont {Lin}}, \bibinfo
  {author} {\bibfnamefont {H.}~\bibnamefont {Liu}}, \bibinfo {author}
  {\bibfnamefont {P.}~\bibnamefont {Yu}}, \bibinfo {author} {\bibfnamefont
  {P.}~\bibnamefont {Tang}}, \bibinfo {author} {\bibfnamefont {S.}~\bibnamefont
  {Meng}}, \bibinfo {author} {\bibfnamefont {W.}~\bibnamefont {Duan}}, \ and\
  \bibinfo {author} {\bibfnamefont {S.}~\bibnamefont {Zhou}},\ }\bibfield
  {title} {\emph {\enquote {\bibinfo {title} {Pseudospin-selective Floquet band
  engineering in black phosphorus},}\ }}\href {\doibase
  10.1038/s41586-022-05610-3} {\bibfield  {journal} {\bibinfo  {journal}
  {Nature}\ }\textbf {\bibinfo {volume} {614}},\ \bibinfo {pages} {75}
  (\bibinfo {year} {2023})}\BibitemShut {NoStop}%
\bibitem [{\citenamefont {Roberts}\ \emph {et~al.}(2011)\citenamefont
  {Roberts}, \citenamefont {Cormode}, \citenamefont {Reynolds}, \citenamefont
  {Newhouse-Illige}, \citenamefont {LeRoy},\ and\ \citenamefont
  {Sandhu}}]{RobertsAPL2011}%
  \BibitemOpen
  \bibfield  {author} {\bibinfo {author} {\bibfnamefont {A.}~\bibnamefont
  {Roberts}}, \bibinfo {author} {\bibfnamefont {D.}~\bibnamefont {Cormode}},
  \bibinfo {author} {\bibfnamefont {C.}~\bibnamefont {Reynolds}}, \bibinfo
  {author} {\bibfnamefont {T.}~\bibnamefont {Newhouse-Illige}}, \bibinfo
  {author} {\bibfnamefont {B.~J.}\ \bibnamefont {LeRoy}}, \ and\ \bibinfo
  {author} {\bibfnamefont {A.~S.}\ \bibnamefont {Sandhu}},\ }\bibfield  {title}
  {\emph {\enquote {\bibinfo {title} {Response of graphene to femtosecond
  high-intensity laser irradiation},}\ }}\href {\doibase 10.1063/1.3623760}
  {\bibfield  {journal} {\bibinfo  {journal} {Appl. Phys. Lett.}\ }\textbf
  {\bibinfo {volume} {99}},\ \bibinfo {pages} {051912} (\bibinfo {year}
  {2011})}\BibitemShut {NoStop}%
\bibitem [{\citenamefont {Ghosh}\ \emph {et~al.}(2026)\citenamefont {Ghosh},
  \citenamefont {Marsal},\ and\ \citenamefont
  {Black-Schaffer}}]{ghosh_2026_18412848}%
  \BibitemOpen
  \bibfield  {author} {\bibinfo {author} {\bibfnamefont {A.~K.}\ \bibnamefont
  {Ghosh}}, \bibinfo {author} {\bibfnamefont {Q.}~\bibnamefont {Marsal}}, \
  and\ \bibinfo {author} {\bibfnamefont {A.}~\bibnamefont {Black-Schaffer}},\
  }\href {\doibase 10.5281/zenodo.18412848} {\enquote {\bibinfo {title}
  {Laser-induced topological phases in monolayer amorphous carbon},}\ }
  (\bibinfo {year} {2026})\BibitemShut {NoStop}%
\bibitem [{\citenamefont {Groth}\ \emph {et~al.}(2014)\citenamefont {Groth},
  \citenamefont {Wimmer}, \citenamefont {Akhmerov},\ and\ \citenamefont
  {Waintal}}]{Groth_2014}%
  \BibitemOpen
  \bibfield  {author} {\bibinfo {author} {\bibfnamefont {C.~W.}\ \bibnamefont
  {Groth}}, \bibinfo {author} {\bibfnamefont {M.}~\bibnamefont {Wimmer}},
  \bibinfo {author} {\bibfnamefont {A.~R.}\ \bibnamefont {Akhmerov}}, \ and\
  \bibinfo {author} {\bibfnamefont {X.}~\bibnamefont {Waintal}},\ }\bibfield
  {title} {\emph {\enquote {\bibinfo {title} {Kwant: a software package for
  quantum transport},}\ }}\href {\doibase 10.1088/1367-2630/16/6/063065}
  {\bibfield  {journal} {\bibinfo  {journal} {New J. Phys.}\ }\textbf {\bibinfo
  {volume} {16}},\ \bibinfo {pages} {063065} (\bibinfo {year}
  {2014})}\BibitemShut {NoStop}%
\bibitem [{\citenamefont {Chadha}\ \emph {et~al.}(2024)\citenamefont {Chadha},
  \citenamefont {Moghaddam}, \citenamefont {van~den Brink},\ and\ \citenamefont
  {Fulga}}]{ChadhaPRB2024}%
  \BibitemOpen
  \bibfield  {author} {\bibinfo {author} {\bibfnamefont {N.}~\bibnamefont
  {Chadha}}, \bibinfo {author} {\bibfnamefont {A.~G.}\ \bibnamefont
  {Moghaddam}}, \bibinfo {author} {\bibfnamefont {J.}~\bibnamefont {van~den
  Brink}}, \ and\ \bibinfo {author} {\bibfnamefont {C.}~\bibnamefont {Fulga}},\
  }\bibfield  {title} {\emph {\enquote {\bibinfo {title} {Real-space
  topological localizer index to fully characterize the dislocation skin
  effect},}\ }}\href {\doibase 10.1103/PhysRevB.109.035425} {\bibfield
  {journal} {\bibinfo  {journal} {Phys. Rev. B}\ }\textbf {\bibinfo {volume}
  {109}},\ \bibinfo {pages} {035425} (\bibinfo {year} {2024})}\BibitemShut
  {NoStop}%
\bibitem [{\citenamefont {Bueno}\ \emph {et~al.}(2007)\citenamefont {Bueno},
  \citenamefont {Furtado},\ and\ \citenamefont {Johnson}}]{BuenoLAA2007}%
  \BibitemOpen
  \bibfield  {author} {\bibinfo {author} {\bibfnamefont {M.}~\bibnamefont
  {Bueno}}, \bibinfo {author} {\bibfnamefont {S.}~\bibnamefont {Furtado}}, \
  and\ \bibinfo {author} {\bibfnamefont {C.}~\bibnamefont {Johnson}},\
  }\bibfield  {title} {\emph {\enquote {\bibinfo {title} {Congruence of
  Hermitian matrices by Hermitian matrices},}\ }}\href {\doibase
  https://doi.org/10.1016/j.laa.2007.03.016} {\bibfield  {journal} {\bibinfo
  {journal} {Linear Algebra Appl.}\ }\textbf {\bibinfo {volume} {425}},\
  \bibinfo {pages} {63} (\bibinfo {year} {2007})}\BibitemShut {NoStop}%
\bibitem [{\citenamefont {Cerjan}\ \emph {et~al.}(2023)\citenamefont {Cerjan},
  \citenamefont {Loring},\ and\ \citenamefont {Vides}}]{CerjanJMP2023}%
  \BibitemOpen
  \bibfield  {author} {\bibinfo {author} {\bibfnamefont {A.}~\bibnamefont
  {Cerjan}}, \bibinfo {author} {\bibfnamefont {T.~A.}\ \bibnamefont {Loring}},
  \ and\ \bibinfo {author} {\bibfnamefont {F.}~\bibnamefont {Vides}},\
  }\bibfield  {title} {\emph {\enquote {\bibinfo {title} {{Quadratic
  pseudospectrum for identifying localized states}},}\ }}\href {\doibase
  10.1063/5.0098336} {\bibfield  {journal} {\bibinfo  {journal} {J. Math.
  Phys.}\ }\textbf {\bibinfo {volume} {64}},\ \bibinfo {pages} {023501}
  (\bibinfo {year} {2023})}\BibitemShut {NoStop}%
\end{thebibliography}%
%============End of MAIN PAPER=============

%==============SUPPLEMENT=============
%\clearpage
\begin{onecolumngrid}
\vspace{1.5cm}
\begin{center}
{\fontsize{12}{12}\selectfont
	\textbf{Supplemental Material for ``Laser-induced topological phases in monolayer amorphous carbon''\\[5mm]}}
 {\normalsize Arnob Kumar Ghosh~\orcidlink{0000-0003-0990-8341}, Quentin Marsal~\orcidlink{0000-0001-6506-1582}, and Annica M. Black-Schaffer~\orcidlink{0000-0002-4726-5247}\\[1mm]}
 {\small \textit{Department of Physics and Astronomy, Uppsala University, Box 516, 75120 Uppsala, Sweden}\\[0.5mm]}
\end{center}
%\tableofcontents 
\vspace{-0.1cm}
\normalsize

%--------------------SM abstract--------------------
%\begin{center}
%\parbox{16cm}{In this supplemental material,} 
%\end{center}
%--------------------SM abstract--------------------

\newcounter{defcounter}
\setcounter{defcounter}{0}
\setcounter{equation}{0}
\renewcommand{\theequation}{S\arabic{equation}}
\setcounter{figure}{0}
\renewcommand{\thefigure}{S\arabic{figure}}
\setcounter{page}{1}
\pagenumbering{roman}
\setcounter{section}{0}
\renewcommand{\thesection}{\arabic{section}}

%------------------------------------------------------
\section{LDLT decomposition}
%------------------------------------------------------
%\textit{LDLT decomposition.---}
In the main text, we discuss the space- and energy-resolved Chern number $C_{x,y,E}$, see Eq.~(3), which involves taking the signature of the  $2N \times 2N$ matrix $L_{x,y, E} \left(X,Y, H_F\right)$, with $N$ being the number of lattice sites in the system. This easily becomes numerically expensive. The computations can be made more numerically efficient by noting that we can write $L_{x,y, E} \left(X,Y, H_F\right)=P_{L} D P_{U}$,  with $P_{L}~(P_U)$ being a block lower~(upper) triangular matrix and $D$ a block diagonal matrix given by ~\cite{ChadhaPRB2024,GhoshSL2024}
\begin{align}
    P_L = 
    \begin{pmatrix}
        \mathbb{I} & \vect{0} \\
        \left(\tilde{X}+ i \tilde{Y} \right) \tilde{H}_F^{-1} & \mathbb{I}
    \end{pmatrix} , \quad
    P_U= 
    \begin{pmatrix}
        \mathbb{I} & \tilde{H}_F^{-1} \left(\tilde{X}- i \tilde{Y} \right)  \\
        \vect{0} & \mathbb{I}
    \end{pmatrix} , \quad
    D = 
    \begin{pmatrix}
        \tilde{H}_F   & \vect{0} \\
       \vect{0} & -\tilde{H}_F - \left(\tilde{X}+ i \tilde{Y} \right) \tilde{H}_F^{-1} \left(\tilde{X}- i \tilde{Y} \right) 
    \end{pmatrix} ,
\end{align}
where we defined $\tilde{X}=(X\! -\!x I)$, $\tilde{Y}=(Y\! -\!y I)$, and $\tilde{H}_F=(H_F\! -\!E I)$. Here, $P_U=P_L^\dagger$ and thus this is equivalent to LDLT factorization, with $L_{x,y, E} \left(X,Y, H_F\right)=P D P^\dagger$, with $P=P_L$. This property allows us to use Sylvester’s law of inertia to obtain ${\rm sig}[L_{x,y, E} \left(X,Y, H_F\right)]={\rm sig}[D]$~\cite{BuenoLAA2007}. Thus, the Chern number $C_{x,y,E}$ can numerically be computed as
\begin{align}
    C_{x,y,E}=\frac{1}{2} \left( {\rm sig} \left[\tilde{H}_F \right] -{\rm sig} \left[ \tilde{H}_F + \left(\tilde{X}+ i \tilde{Y} \right) \tilde{H}_F^{-1} \left(\tilde{X}- i \tilde{Y} \right) \right] \right) \ .
    \label{chernNoLDLT}
\end{align}
Thus, instead of diagonalizing the full spectral localizer matrix, we only diagonalize lower-dimensional matrices: $\tilde{H}_F$ and $\tilde{H}_F + \left(\tilde{X}+ i \tilde{Y} \right) \tilde{H}_F^{-1} \left(\tilde{X}- i \tilde{Y} \right)$. We use Eq.~\eqref{chernNoLDLT} to compute the spatial averaged $C_{0,\pi}$ and configuration averaged $\bar{C}_{0,\pi}$ Chern numbers,  making the computations much faster.

%------------------------------------------------------
\section{Localizer gap}
%------------------------------------------------------
% \textit{Localizer gap.---}
Beyond the extracted Chern number from the spectral localizer in Eq.~(2) in the main text, 
the localizer gap additionally provides important information, in particular, the location of the edge states. The localizer gap $\sigma_{\rm L}^{x,y,E}\left(X,Y,H_F\right)$ is defined as~\cite{CerjanJMP2023}
\begin{align}
    \sigma_{\rm L}^{x,y,E}\left(X,Y,H_F\right)= {\rm min} \left( \lvert \sigma \left[L_{x,y,E}(X,Y,H_F) \right] \rvert \right),
    \label{eq:Lgap}
\end{align}
where $\sigma \left[L_{x,y,E}(X,Y,H_F) \right]$ represents the spectrum of spectral localizier matrix $L_{x,y,E}(X,Y,H_F)$. The localizer gap $ \sigma_{\rm L}^{x,y,E}\left(X,Y,H_F\right)$ vanishes if there exists a boundary or edge state at the position $x,y$ and at the energy $E$, and thereby acts as a local topological band gap. The localizer gap thus remains finite inside the bulk of the system and outside of the system, but goes to zero at the edges of the system. In order to aid plotting, we consider the normalized localizer gap $\sigma_{\rm N}^{x,y,E}\left(X,Y,H_F\right)=\sigma_{\rm L}^{x,y,E}\left(X,Y,H_F\right)/{\rm max}\left[\sigma_{\rm L}^{x,y,E}\left(X,Y,H_F\right)\right]$. For brevity, we exclude superscripts and arguments for the normalized localizer gap and denote it simply as $\sigma_{\rm N}$ only.

%------------------------------------------------------
\section{Spatially resolved localizer gap and topological invariant}
%------------------------------------------------------
% \textit{Spatially resolved localizer gap and topological invariant.---}
In Fig.~2 in the main text, we discuss the eigenvalue spectra and local density of states~(LDOS), which illustrate the signatures of the $0$- and $\pi$-modes. In Fig.~\ref{Fig:SM2}, we provide complementary data in the form of the spatially resolved normalized localizer gap $\sigma_{\rm N}$ and the topological invariants of the $0$- and $\pi$-modes $C_{x,y,E={0,\pi}}$.
Focusing first on driven crystalline graphene in Figs.~\ref{Fig:SM2}(a-d), the localizer gap $\sigma_{\rm N}$ vanishes only near the edges of the system, while it is finite both inside and outside of the system, signifying an effective local band gap in the system. The Chern number $C_{x,y,E=0}$ follows the behavior of the localizer gap and takes a finite uniform value $C_0=-1$ inside the bulk of the system but vanishes near and outside the system's edge. For $E=\pi$, we observe a similar behavior of the localizer gap and Chern number $C_\pi$, as shown in Figs.~\ref{Fig:SM2}(c,d). For $C_{x,y,E=\pi}$, we observe some variation near the edge of the system, but it takes a stable, uniform value of $-4$ inside the bulk of the system. 

Next, we consider driven monolayer amorphous carbon. Figures~\ref{Fig:SM2}(e,f) illustrate the localizer gap and Chern number computed at $E=0$. Both the localizer gap and the Chern number exhibit a very similar behavior to crystalline graphene, always carrying a stable Chern number, signifying the topological nature of driven monolayer amorphous carbon. This result is also not dependent on the amount of non-hexagons in the amorphous lattice.
However, for $E=\pi$ in Figs.~\ref{Fig:SM2}(g,h), the localizer gap $\sigma_{\rm N}$ shows multiple closings ($\sigma_{\rm N}=0$) inside the bulk of the system. This behavior of $\sigma_{\rm N}$, is mimicked in the Chern number $C_{x,y,E=\pi}$ which switches between multiple non-zero values inside the system. As a consequence, the spatially averaged $C_\pi$ gets a finite variance as shown in Fig.~3(d) in the main text. Also, the spatially varying behavior of the $C_{x,y,E=\pi}$ is also responsible for the delocalization of its associated edge state as seen in the LDOS plot in Fig.~2(f). We note that the computation of the topological invariant in the $E=\pi$ gap may be affected by finite-size effects, and in the thermodynamic limit, it may be possible to recover a stable topological phase. 

%~~~~~~~~~~~~~~~~~~~~~~~~~~~~~~~~~~~~~~~~~~~~~~~~~~~~~~~~~
%~~~~~~~~~~~~~~~~~~~~~~~~~~~~~~~~~~~~~~~~~~~~~~~~~~~~~~~~~
\begin{figure}
	\centering
	\subfigure{\includegraphics[width=0.81\textwidth]{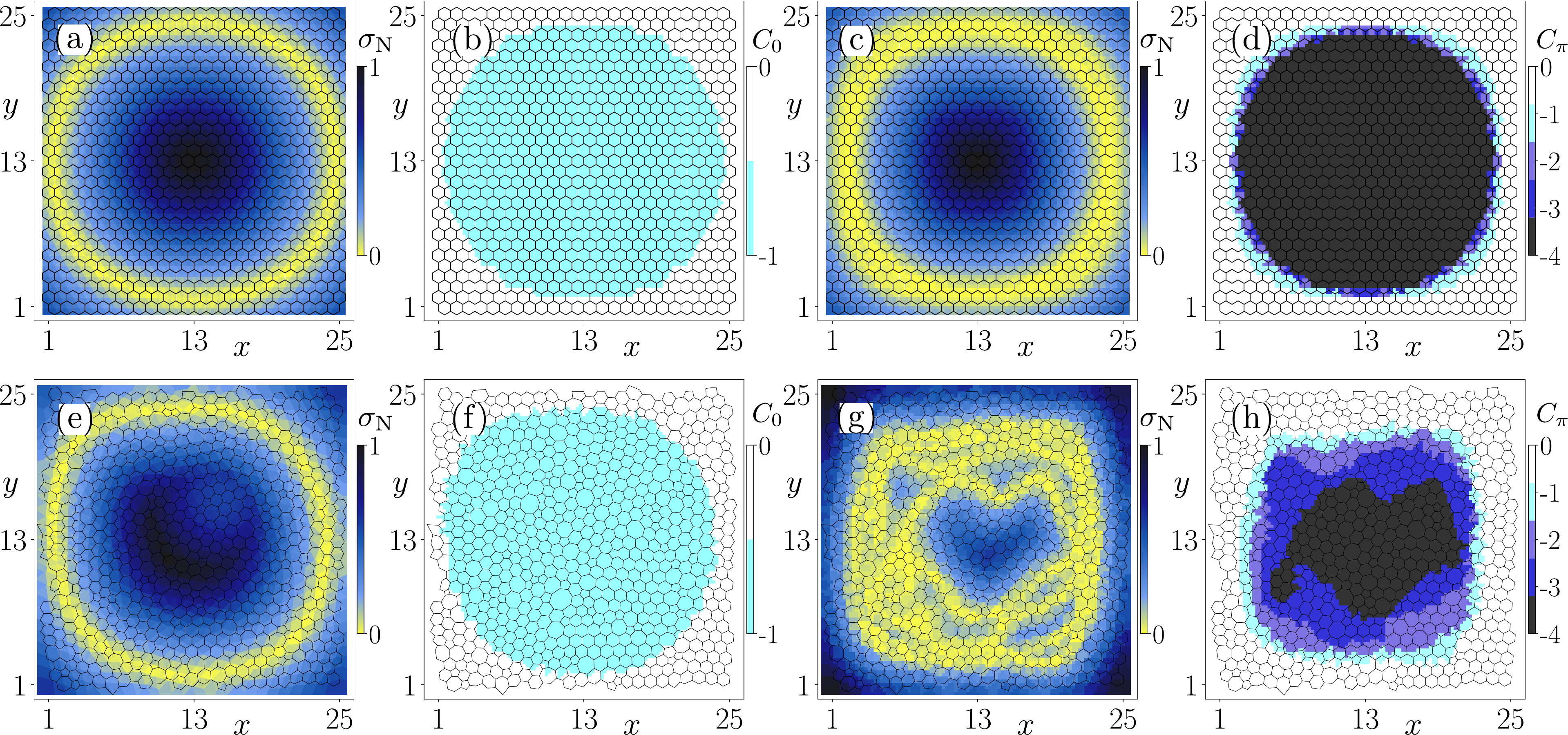}}
	\caption{(a)~[(c)] Spatially resolved normalized localizer gap $\sigma_{\rm N}$ and  (b)~[(d)] Chern number $C_{x,y,E=0}$~[$C_{x,y,E=\pi}$] computed at $E=0$~[$\pi$] for driven crystalline graphene.  (e-h) Repeats (a-d), but for driven monolayer amorphous carbon. Parameters are the same as Fig.~2 in the main text.
	}
	\label{Fig:SM2}
\end{figure}
%~~~~~~~~~~~~~~~~~~~~~~~~~~~~~~~~~~~~~~~~~~~~~~~~~~~~~~~~~
%~~~~~~~~~~~~~~~~~~~~~~~~~~~~~~~~~~~~~~~~~~~~~~~~~~~~~~~~~

%------------------------------------------------------
\section{Stable $\pi$-mode}
%------------------------------------------------------
% \textit{Stable $\pi$-mode.---}
The results in the main text show that the $\pi$-modes are largely unstable in driven monolayer amorphous carbon. However, there exists a small parameter regime where the Chern number variance vanishes in Fig.~3(d), indicating the possibility of finding stable $\pi$-modes also in monolayer amorphous graphene. In Fig.~\ref{Fig:SM2-2}, we demonstrate one such choice of parameters for the drive hosting a stable $\pi$-mode. 
We plot the eigenvalue spectrum $E_m$ as a function of eigenvalue index $m$ in Fig.~\ref{Fig:SM2-2}(a), color coding the eigenvalues with the IPR of that state. Here, we only focus on the modes appearing in the $\pi$-gap and show the zoomed-in spectrum in the inset, which indicates the existence of localized states in the $\pi$-gap. Subsequently, we plot the LDOS computed inside the $\pi$-gap in Fig.~\ref{Fig:SM2-2}(b). The LDOS reveals the presence of states near the edge of the system, although there are also some finite states in the bulk of the system. Nevertheless, the $\pi$-mode that appears in this case has a more prominent edge presence than in Figs.~2(d-f) in the main text. To further investigate these modes, we compute the spatially resolved normalized localizer gap $\sigma_{\rm N}$ and Chern number $C_{x,y,E=\pi}$ in Figs.~\ref{Fig:SM2-2}(c,d). Here, the localizer gap is more stable than that of Fig.~\ref{Fig:SM2-2}(g). The corresponding Chern number $C_\pi$ is also mostly stable inside the bulk of the system, apart from some fluctuations near the boundaries. These results establish that monolayer amorphous carbon can support a stable $\pi$-mode if the drive is tuned appropriately.
%~~~~~~~~~~~~~~~~~~~~~~~~~~~~~~~~~~~~~~~~~~~~~~~~~~~~~~~~~
%~~~~~~~~~~~~~~~~~~~~~~~~~~~~~~~~~~~~~~~~~~~~~~~~~~~~~~~~~
\begin{figure}
	\centering
	\subfigure{\includegraphics[width=0.85\textwidth]{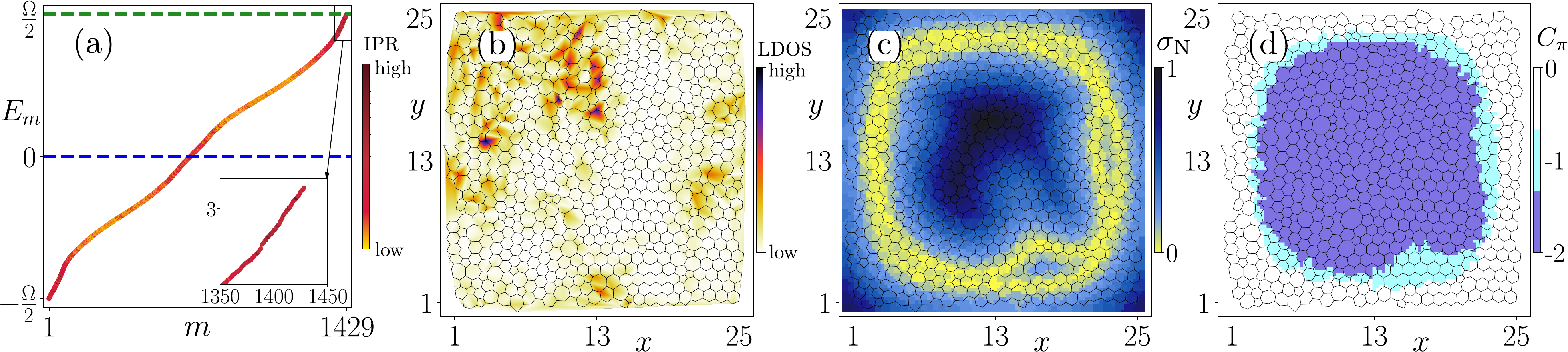}}
	\caption{(a) Quasienergy spectrum $E_m$ as a function of eigenstate index $m$ for driven monolayer amorphous carbon. Color encodes the IPR of the given state. Inset shows zoomed-in spectrum close to $E=\pi$. (b) Spatially resolved LDOS associated with states at quasienergy gap $\pi$. (c,d) Spatially resolved localizer gap $\sigma_{\rm N}$ and Chern number $C_{x,y,E=\pi}$ computed at $E=\pi$. Here, $A=2.9, \Omega=1.2$, using the same amorphous lattice as in Fig.~2 in the main text.
	}
	\label{Fig:SM2-2}
\end{figure}
%~~~~~~~~~~~~~~~~~~~~~~~~~~~~~~~~~~~~~~~~~~~~~~~~~~~~~~~~~
%~~~~~~~~~~~~~~~~~~~~~~~~~~~~~~~~~~~~~~~~~~~~~~~~~~~~~~~~~

%------------------------------------------------------
\section{Variance}
%------------------------------------------------------
In the main text, we use a non-zero variance of the extracted Chern number to indicate a relative instability of the topological phase, since a stable topological phase should have its topological invariant remaining uniform throughout the system's bulk. 
The variance is defined as $\sigma^2_E=\frac{1}{N_P}\sum_{i=1}^N \left(C_{i,E} -C_E\right)^2$, with $N_P$ the number of points considered inside the bulk of the system to compute the Chern number, $C_{i,E}$ the local Chern number at energy gap $E$ at lattice site $i$, and $C_E$ is the average Chern number. 
Furthermore, we take an average of $\sigma^2_E$ over disorder configurations and denote that as $\bar{\sigma}^2_E$. 
We plot $\bar{\sigma}^2_E$ in the $A\mhyphen \Omega$ plane in Fig.~\ref{Fig:Variance}(a) and (b) for the $0$ and $\pi$-gap, respectively, using the same setup as in Fig.~3(c,d) in the main text. 
The variance is finite in the low-frequency regime for both $0$ and $\pi$ gaps. 
It is also finite close to bulk gap closings, as expected due to a small spectral gap in these regions. 
However, for the $0$-gap, there still exists a large region with a zero variance and non-trivial topology  [see Fig.~3(c) in the main text], showcasing stable topological modes. 
In contrast, for the $\pi$-gap, the region with finite topology [see Fig.~3(d) in the main text] and vanishing variance is very small. 
We attribute this reduced stability of the $\pi$-modes to a small spectral gap protecting these modes.

%~~~~~~~~~~~~~~~~~~~~~~~~~~~~~~~~~~~~~~~~~~~~~~~~~~~~~~~~~
%~~~~~~~~~~~~~~~~~~~~~~~~~~~~~~~~~~~~~~~~~~~~~~~~~~~~~~~~~
\begin{figure}[h!]
	\centering
	\subfigure{\includegraphics[width=0.55\textwidth]{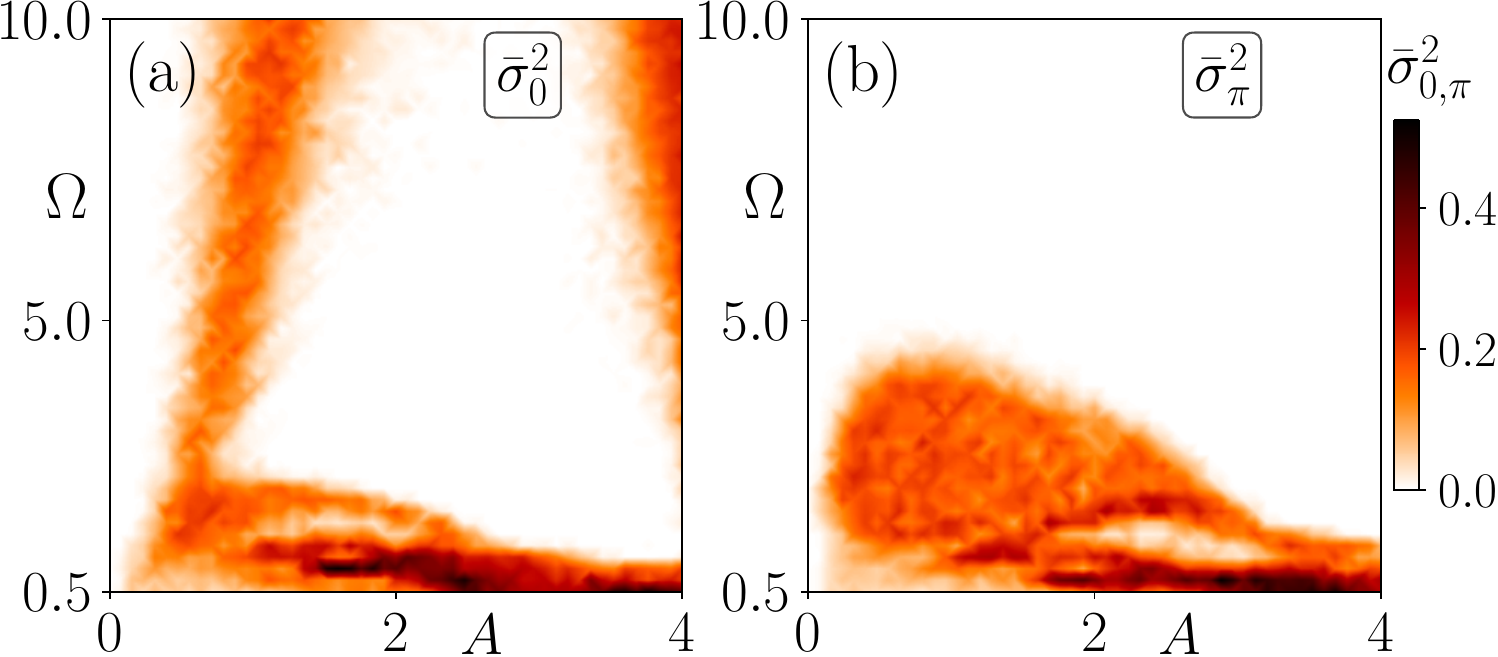}}
	\caption{Variance $\bar{\sigma}^2_E$ in the $A \mhyphen \Omega$ plane for the (a) $0$-gap (b) $\pi$-gap. Same setup as in Fig.~3(c,d) in the main text.
	}
	\label{Fig:Variance}
\end{figure}
%~~~~~~~~~~~~~~~~~~~~~~~~~~~~~~~~~~~~~~~~~~~~~~~~~~~~~~~~~
%~~~~~~~~~~~~~~~~~~~~~~~~~~~~~~~~~~~~~~~~~~~~~~~~~~~~~~~~~

\end{onecolumngrid}

\end{document}